\RequirePackage{lineno}
\documentclass[aps, prl,
floatfix,
superscriptaddress,showpacs]{revtex4-1}
\usepackage{t1enc}
\usepackage{hyperref}
\usepackage[english]{babel}
\usepackage{graphicx}
\usepackage{epsfig}
\usepackage{xspace}
\usepackage{natbib}

\newcommand{\smhb}{\ensuremath{\mathrm{SmB}_6}\xspace}
\newcommand{\Xpoint}{\ensuremath{X}-point\xspace}
\providecommand{\etal}{et al.\xspace}
\providecommand{\ie}{i.e.,\xspace}
\providecommand{\eg}{e.g.,\xspace}

\providecommand{\Gbar}{\ensuremath{\bar{\Gamma}}\xspace}
\providecommand{\fourf}{4$f$\xspace}

\providecommand{\Xbar}{\ensuremath{\bar{X}}\xspace}
\providecommand{\Mbar}{\ensuremath{\bar{M}}\xspace}
\providecommand{\Hbar}{\ensuremath{\bar{H}}\xspace}
\providecommand{\ppol}{$p$-polarization\xspace}
\providecommand{\spol}{$s$-polarization\xspace}
\providecommand{\twodim}{two-dimen\-sional\xspace}
\providecommand{\threedim}{three-dimen\-sional\xspace}
\providecommand{\fmult}{$f$-orbital multiplet\xspace}
\providecommand{\fmults}{$f$-orbital multiplets\xspace}
\providecommand{\sHfivetwo}{\ensuremath{^6H_{5/2}}\xspace}
\providecommand{\flevel}{$f$-level\xspace}
\providecommand{\hv}[1]{$h\nu = #1$~eV\xspace}
\providecommand{\dband}{$d$-band\xspace}
\providecommand{\Bos}{B~1$s$\xspace}
\providecommand{\kpar}{\ensuremath{k_{||}}\xspace}
\providecommand{\smtwopl}{Sm$^{2+}$\xspace}
\providecommand{\smthreepl}{Sm$^{3+}$\xspace}
\providecommand{\kperp}{$k_\perp$\xspace}
\providecommand{\rshbpr}{\ensuremath{\alpha_R}\xspace}

\providecommand{\tss}{topological surface state\xspace}
\providecommand{\tsss}{topological surface states\xspace}

\providecommand{\mdcs}{momentum distribution curves\xspace}
\providecommand{\be}{binding energy\xspace}
\providecommand{\bes}{binding energies\xspace}
\providecommand{\eb}{\ensuremath{E_B}\xspace}
\providecommand{\scls}{surface core level shift\xspace}

\begin{document}

\title{Samarium hexaboride: A trivial surface conductor}

\author{P. Hlawenka}
\author{K. Siemensmeyer}
\author{E. Weschke}
\author{A. Varykhalov}
\author{J. S\'anchez-Barriga}
\affiliation{Helmholtz-Zentrum Berlin f\"ur Materialien und
Energie, Elektronenspeicherring BESSY II, Albert-Einstein-Stra\ss
e 15, 12489 Berlin, Germany}

\author{N. Y. Shitsevalova}
\author{A. V. Dukhnenko}
\author{V. B. Filipov}
\affiliation{Institute for Problems of Materials Science,
National Academy of Sciences of Ukraine, Krzhyzhanovsky str. 3,
03142 Kiev, Ukraine}

\author{S. Gab\'ani}
\author{K. Flachbart}
\affiliation{Institute of Experimental Physics, Slovak Academy
of Sciences, Watsonova 47, 04001 Ko\v sice, Slovakia}

\author{O. Rader}
\author{E. D. L. Rienks}
\affiliation{Helmholtz-Zentrum Berlin f\"ur Materialien und Energie, Elektronenspeicherring BESSY II, Albert-Einstein-Stra\ss e 15,  12489 Berlin, Germany}

\date{\today}

\begin{abstract}

Recent theoretical and experimental studies suggest that
\smhb is the first topological Kondo insulator: A material in which
the interaction between localized and itinerant electrons renders
the bulk insulating at low temperature, while topological surface
states leave the surface metallic.
While this would elegantly explain the material's puzzling conductivity,
we find the experimentally observed candidates for both predicted
topological surface states to be of trivial character instead: The
surface state at \Gbar is very heavy and shallow with
a mere $\sim  2$ meV \be. It exhibits large Rashba splitting which
excludes a topological nature. We further demonstrate that the other
metallic surface state, located at \Xbar, is not an independent
in-gap state as supposed previously, but part of a massive band
with much higher \be ($ 1.7$ eV). We show that it remains metallic
down to 1 K due to reduced hybridization with the energy-shifted
surface \fourf level.
\end{abstract}


\maketitle

\section{Introduction}

Over the past decade, topological insulators have been established
as a new state of matter \cite{Hasan:2010ku,*Bernevig:un}.  
Rather subtle symmetry attributes of
the bulk band structure of these materials dictate the electronic
structure at their surface: Gapless states with a particular
spin polarization will inevitably exist at the confines of insulators
that can be classified as topologically non-trivial.

The about 30 materials that have thus far been positively identified
as members of this category \cite{Ando:2013fg} are well captured
by a single electron description. In recent years, the focus has
shifted towards the question whether the concept of topological
insulators can be generalized to materials in which electron
correlation plays a more prominent role. Two families of materials
with increased correlation strength have been brought forward as
promising candidates: Ir-based materials
\cite{Shitade:2009he,*Guo:2009gj} and the Kondo
insulators \cite{Dzero10}. In the latter group, hybridization between
a narrow \flevel and the much wider conduction band leads to a small
indirect band gap at low temperature. At appropriate band filling,
the chemical potential falls within this gap, rendering the material
an insulator. In the case of \smhb , the suggestion that it could
be a topological insulator
\cite{Dzero10,Takimoto11,LuF13,Alexandrov13,Roy14,Neupane13,JiangNC13,XuPRB13,XuNC14,Min14,Denlinger13126637,LiG14}
has deservedly attracted a lot of attention, since it would also
provide an elegant solution to the long-standing controversy about
its unexpected residual conductivity at low temperature
\cite{Cooley95,Kikoin95,Kasuya96,Riseborough00,Sluchanko00}. Could
\smhb , after being identified as the first mixed valence system
in the 1960s \cite{Vainshtein65} and the first Kondo insulator
shortly after \cite{Menth69,Nickerson71,Allen79}, now become the
exemplification of yet another phenomenon in solid state research?

Band structure calculations, treating correlation in various ways,
do point in this direction
\cite{Takimoto11,LuF13,Alexandrov13,Roy14}. A $d$-$f$ band
inversion at $X$ is predicted to give rise to two distinct topological
surface states: One at \Xbar, appearing twice in the surface Brillouin
zone, and the second at \Gbar . Three Dirac cones are thus predicted,
which would make \smhb a strong topological insulator. The cubic
symmetry is further found to exclude it from having a phase with
an even number of Dirac cones, \ie from being a weak topological
insulator \cite{Alexandrov13}.

The first strides on the experimental front were also seemingly
successful towards establishing \smhb as the first topological Kondo
insulator: It is indeed found that it behaves like a bulk insulator
with a conductive surface at low temperature
\cite{Wolgast:2013ih,ZhangXPRX13,KimDJSmGdB6}. In addition, evidence
of metallic surface states at the predicted sites, \Gbar and \Xbar
, has been found using angle-resolved photoemission spectroscopy
(ARPES)
\cite{Neupane13,JiangNC13,XuPRB13,Frantzeskakis13,XuNC14,Min14,Denlinger13126637}
and de Haas-van Alphen experiments \cite{LiG14}.

Even though several aspects leave ample room for doubts, most notably
the fact that no Dirac cones have been found and the much lower
than expected effective mass \cite{Alexandrov13}, two reports have appeared
that provide seemingly strong support in favor of the topological
thesis: It is found that the introduction of magnetic impurities
suppresses the residual low-temperature conductance \cite{KimDJSmGdB6}.
Moreover, spin-polarized ARPES measurements appear to confirm the
helical spin texture \cite{XuNC14}.

In spite of these results, we provide strong evidence that the
surface states at the (100) surface of \smhb  are topologically
trivial: We first demonstrate that the ---until now elusive--- state
at \Gbar is massive and exhibits Rashba splitting, excluding a
topological nature. This result already leads to insurmountable
disagreement with the theoretical predictions, since a \tss at \Gbar is
required to arrive at an odd number of Dirac cones. In a
second step we show that the well-characterized \Xbar state is not
an independent in-gap state as supposed previously. Instead, we
show that it is part of a massive surface state with a binding
energy (\eb) of $1.7$ eV that derives from the bulk \dband at $X$. We
demonstrate how this state hybridizes with the surface \fourf level
to yield a gap that is both smaller and shifted to higher \be with
respect to its bulk counterpart.

While our findings thus provide an explanation for the long-standing
puzzle of samarium hexaboride's low temperature conductivity, they
conflict very clearly with the idea that it would be the first
realization of a strongly correlated topological insulator.

\section{The \Gbar state}
 
\begin{figure}
\includegraphics[width=.8\linewidth]{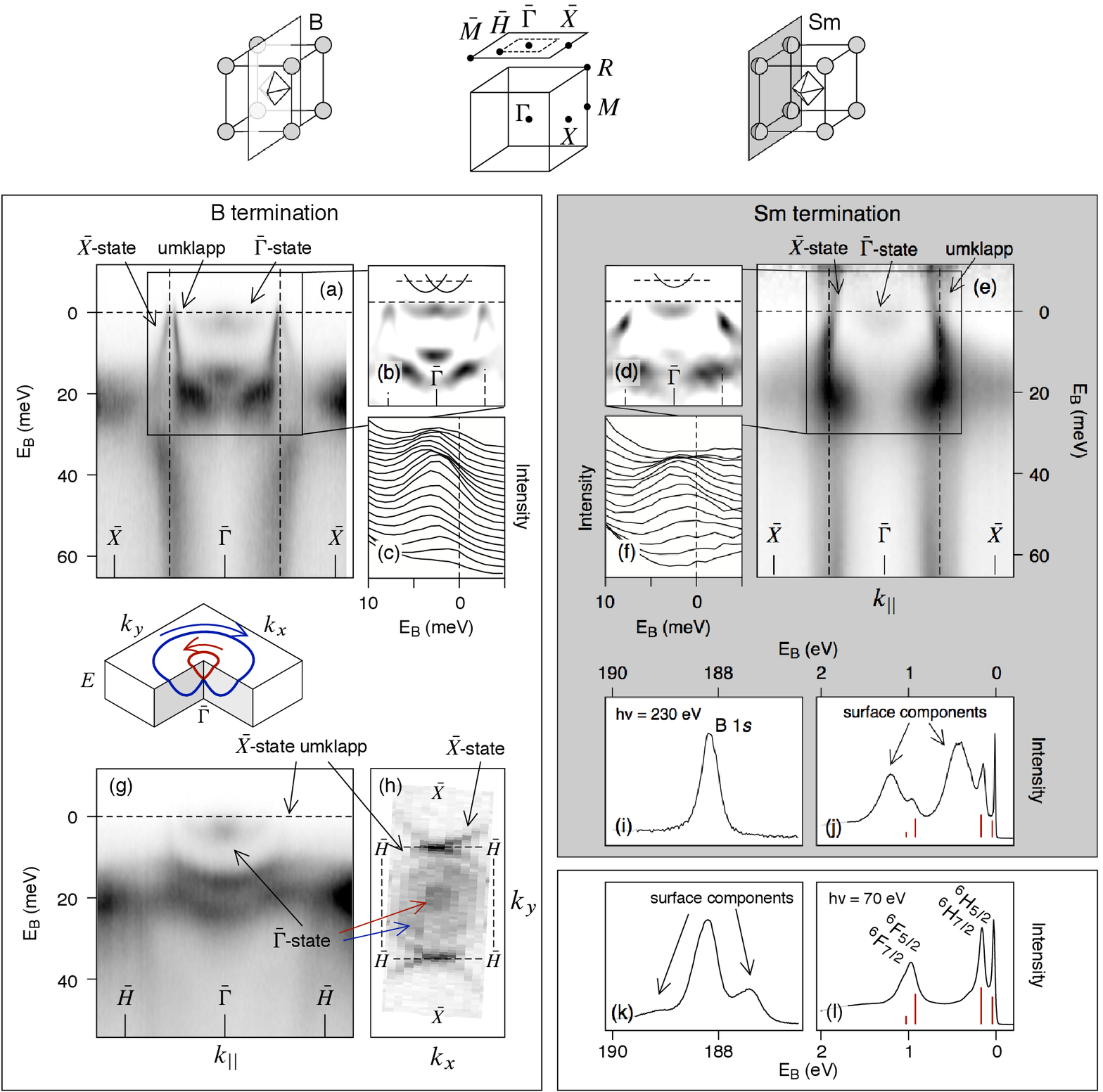}
\caption{Massive nature and Rashba splitting of the \Gbar state.
Photoemission intensity ($I$) along the \Gbar--\Xbar direction for
(a)  B and (e) Sm terminated surfaces together with (b,d) second
derivative ($d^2I / d E^2$).  Solid lines in (b) (Rashba split) and (d) (not split) are parabolic
dispersions derived from a fit of the photoemission intensity with
a \twodim model of the spectral function (vertically offset).  (c,f)
\Gbar state represented as spectra. The data on the Sm terminated
sample (d--f) were measured at 40~K to thermally populate a larger
fraction of the shallow \Gbar state. The results have been divided
by a Fermi-Dirac distribution.  (g) Photoemission intensity along
the \Gbar--\Hbar--\Mbar direction (to avoid contributions from the \Xbar state)
 on the B terminated surface.  (h) Fermi
surface of a B terminated sample. (i,k)  \Bos  core level and (j,l)
angle integrated valence band spectra for B and Sm
terminated samples. Bars in (j,l) indicate calculated 
intensity for the $f^6 \rightarrow f^5$ (\smtwopl ) photoemission
transition \cite{Gerken:1983wk}.  \hv{31} and \spol, unless stated otherwise.}
\label{fig:G}
\end{figure}

In agreement with earlier ARPES studies
\cite{Neupane13,JiangNC13,XuPRB13,XuNC14,Min14,Frantzeskakis13,Denlinger13126637},
we observe a state that forms an elliptical Fermi surface contour
around the \Xbar point. In addition, we can clearly resolve a shallow
feature at the center of the surface Brillouin zone which has
previously been considered a candidate for the Dirac cone at \Gbar
\cite{Neupane13,JiangNC13,XuPRB13,XuNC14,Min14,Frantzeskakis13}.
In Fig.~\ref{fig:G} we show data for B and Sm terminated surfaces 
\footnote{We find that samples cleave with two
distinct terminations as is further described in the Methods section.}.
On the B terminated surface, this state has a handlebar moustache-like
dispersion, reminiscent of a shallow, Rashba-split pair, see
Figs.~\ref{fig:G}(a--c).
The Rashba splitting is a spin-orbit effect
\footnote{See \eg
\href{http://iopscience.iop.org/1367-2630/focus/Focus\%20on\%20the\%20Rashba\%20Effect/}{New J. Phys., Focus on the Rashba
Effect.} } which lifts the spin degeneracy of two-dimensional
states leading to an in-plane spin texture as indicated in
Fig.~\ref{fig:G}.
Alternatively, we point out that the observed dispersion also
bears a resemblance with the \tss of Bi$_2$Te$_3$.
In that system the dispersion becomes non-linear immediately below the
Dirac point and passes through two turning points before it meets with
the valence band \cite{Chen:2009do}. One could thus argue that what we
observe is part of the dispersion of a \tss .
We discard this possibility for two reasons. Firstly, we do not
observe any evidence for the dispersion bending down to connect to
the valence band. This also holds when the state is traced along
the \Gbar--\Mbar direction where a downturn in the dispersion would
not be obscured by the intensity of the \Xbar state, Fig.~\ref{fig:G}(g).
The second, more powerful argument is the termination dependence
we observe. The \Gbar state appears as a single parabola on
the Sm terminated surface, as shown in Fig.~\ref{fig:G}(d--f). This result
strongly hints at Rashba splitting, because this effect relies on
the gradient of the crystal potential perpendicular to the surface. Given
the polar nature that bulk truncated surfaces with pure B or Sm
termination would have, a difference in surface potential is very
conceivable for the different terminations.

Interpreting the \Gbar state as a free electron-like state,
we obtain a \be of $2.3 \pm 0.6$ meV and a very large
effective mass of $\sim 35$ $m_e$. The Rashba parameter (\rshbpr )
amounts to ($1.9 \pm 0.2) \times 10^{-12}$ eV~m on the B terminated
surface.
This very modest Rashba parameter further cements the idea
that the splitting is controlled by a difference in surface
potential between terminations: In systems where \rshbpr is an order
of magnitude larger, such as the $L$-gap surface state on
Au(111) with $\rshbpr = 3 \times 10^{-11}$ eV~m, the observed
splitting can still be influenced by changes in the surface
potential~\cite{Nicolay:2001fy}.  In so-called giant Rashba systems
where the Rashba parameter is an order of magnitude higher still
($\rshbpr \sim 10^{-10}$ eV~m), the splitting is found unaffected
even by large work function changes because it is more difficult to
change the wavefunction asymmetry \cite{Bihlmayer06,Varykhalov:2012ec}.
It is thus conceivable that a relatively small change in the surface
potential between terminations suffices to determine whether the \Gbar
state exhibits Rashba splitting or not.
Note that the splitting is only observable here by virtue of the very large
effective mass, since $\Delta k_{||} = m^* \rshbpr / \hbar^2$.
 
The large effective mass directly suggests that the \Gbar
state stems from a bulk \fourf state.  
Moreover, the intensity of the \Gbar state displays discrete maxima in the
vicinity of the photon energies with which a $\Gamma$ point can be
reached and is next to invisible at all other photon energies.
This further indicates that it is split off a bulk $\Gamma$ state. 

In summary, we find evidence for a massive surface state at the
surface Brillouin zone center which can exhibit Rashba splitting.
As we mentioned above, the absence of a \tss at \Gbar
precludes \smhb from classification as a strong topological insulator,
regardless of the nature of the other surface state at \Xbar . In
the following, however, we will also present a trivial explanation
for this widely observed feature.


\section{The \Xbar state} 

\begin{figure}
\centering
\includegraphics[width=0.7\linewidth]{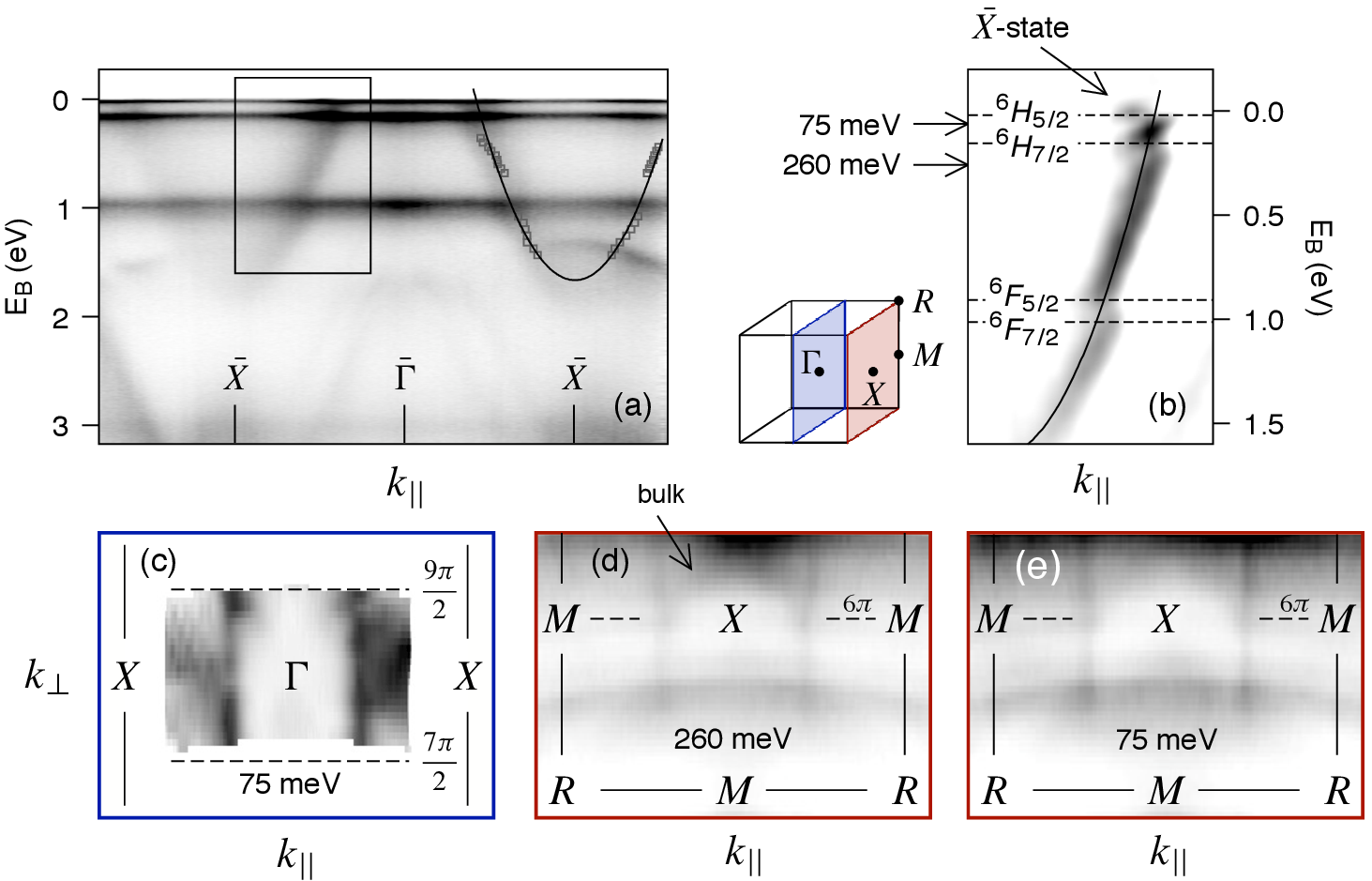}
\caption{Massive, two-dimensional nature of the \Xbar state.
(a) Photoemission intensity along \Xbar--\Gbar--\Xbar obtained with
\hv{70}.  Markers indicate position of maxima in fits of \mdcs ,
solid line is a parabolic fit to these points
(markers and solid line horizontally offset by $2\pi$).  (b) Second
derivative (${d^2I}/{dk^2}$) of the area marked in (a).  Energies
of the \fmults are indicated by dashed lines, arrows indicate the
binding energies of (\kpar, \kperp ) maps shown in (c--e). Photoemission
intensity on the $\Gamma$--$X$--$M$ (blue in the Brillouin zone sketch)
and $X$--$M$--$R$ (red) planes at the indicated binding energies. The
vertical features in (c--e) indicate two-dimensional character. All
results obtained with \ppol .}

\label{fig:X}
\end{figure}

In agreement with published results
\cite{Neupane13,JiangNC13,XuPRB13,XuNC14,Min14,Denlinger13126637,Frantzeskakis13},
we observe a $d$-like band with a \be of about 1.7 eV at $X$, shown
in Figs.~\ref{fig:X}(a) and (b).
In all previous 
work \cite{Neupane13,JiangNC13,XuPRB13,XuNC14,Min14,Denlinger13126637,Frantzeskakis13}, 
this deep state has
been interpreted as the bulk band that hybridizes with the \fmult
to open the bulk gap at low temperature.

While our data are compatible with the existence of such a \threedim
band, we find clear evidence that the prominent parabolic dispersion
observed here and previously
\cite{Neupane13,JiangNC13,XuPRB13,XuNC14,Min14,Denlinger13126637,Frantzeskakis13}
is a surface state: Photon energy dependent measurements unequivocally
show the existence of a \twodim state at \bes well beyond that of
the shallowest \fmult (\sHfivetwo).  The photoemission intensity
on both the $\Gamma$--$X$--$M$ and the $X$--$M$--$R$ planes is shown
in Figs.~\ref{fig:X}(c) and (d--e) respectively.  While traces of
the \threedim \dband around $X$ can be observed ---most prominently
in Fig.~\ref{fig:X}(d)--- the vertical features in all three panels
provide clear evidence of a state that does not display a dependence
with the electron wave vector perpendicular to the surface \kperp.

Denlinger \etal have also attempted to determine the
dimensionality of the electronic structure beyond the \fmult using photon
energy dependent ARPES measurements \cite{Denlinger13126637}. In
contrast to our findings, they do not find evidence of the
\twodim state below the \fmult .  We attribute
this discrepancy to their choice of \be 
that is relatively close to the hybridization region.
We observe features not dispersing with \kperp
at all binding energies sufficiently far away from the multiplets,
where the signal is not dominated by the $f$ intensity.
We further note that the photoemission intensity from the
surface state varies strongly with light polarization and is
suppressed with \spol in the given geometry.

The existence of this deep, massive surface state invalidates
the premise of earlier studies that the elliptical Fermi surface
contour around \Xbar can be attributed to an independent in-gap state. In fact,
due to its group velocity, the massive band at \Xbar will inevitably
persist above the \sHfivetwo \fmult . This provides an alternative, trivial
explanation for the elliptical Fermi surface contour around \Xbar
reported in all ARPES experiments.
In the following we will detail how this metallic feature 
springs from the deep surface state at \Xbar .

\begin{figure}
\centering
\includegraphics[width=.6\linewidth]{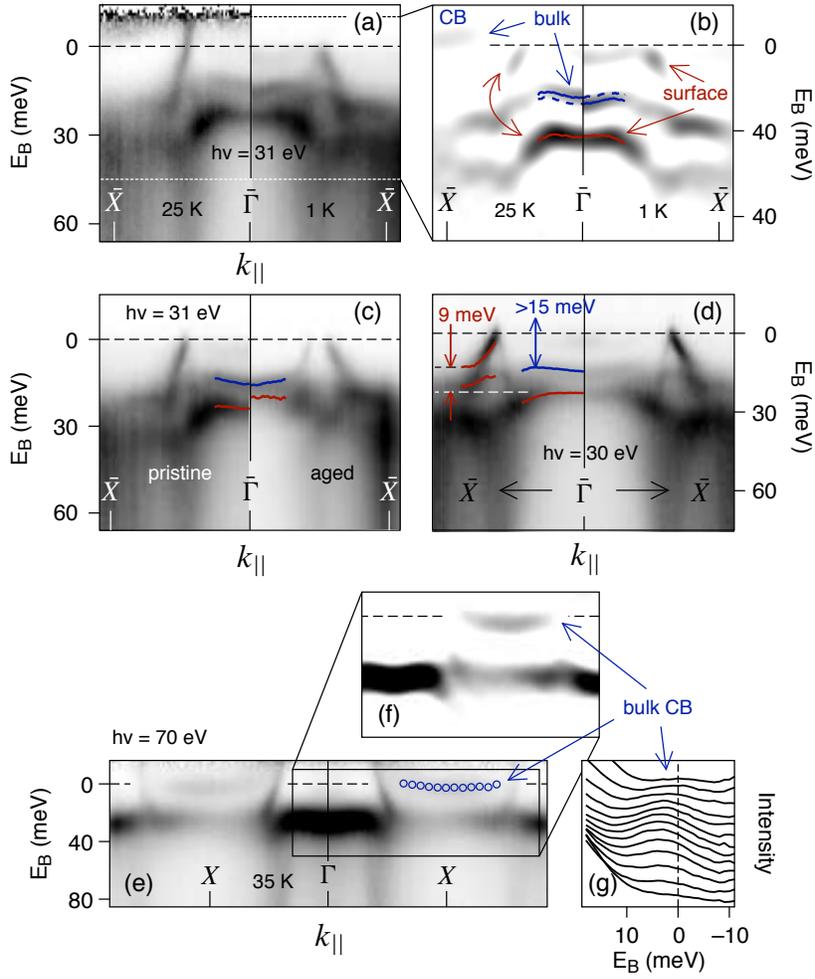}
\caption{Bulk and surface $d$-$f$ hybridization.
(a--c) Evolution of bulk and
surface features as a function of temperature and sample ageing.
(b) Second derivative ($d^2I/dE^2$) of the indicated energy range
of (a).
(e--g) Dispersion of the bulk conduction band (CB) near $X$.
(f) Second derivative ($d^2I/dE^2$) of the indicated area.
(g) Conduction band as spectra.  Photoemission intensity in the
left half of (a) and in (e--g) is divided by the Fermi-Dirac
distribution at the corresponding temperature.  Solid lines in
(b--d) and markers in (e) indicate maxima of energy distribution
curve fits. Dashed lines in (b) indicate the positions of maxima
in the other half of the graph.  Intensity in left and right halves
of (d) and (e) mirrored about $\kpar = 0$. All measurements with
$p$-polarized light.}
\label{fig:hyb}
\end{figure}

In accordance with the results of Min \etal \cite{Min14} and Denlinger \etal
\cite{Denlinger13126637}, we observe how the bulk
conduction band appears to move above the Fermi level upon cooling
below a temperature of about 30 K. This is
illustrated in Fig.~\ref{fig:hyb}(a) and (b).
Simultaneously with the disappearance of the bulk
conduction band at \Xbar we find that the upper maximum of the $f$ multiplet at $\eb \sim 15$
meV around \Gbar , shifts downwards when the temperature is lowered from 25 to 1 K.
The latter band can thus be identified as the bulk valence
band that is pushed to higher \be as the size of the hybridization
gap increases.  We tentatively assign the \flevel with nearly
identical dispersion below it to the surface counterpart of this
band, because it does not show this temperature effect.  This
assignment is corroborated by studying the effect of sample ageing
on the surface electronic structure, detailed in Fig.~\ref{fig:hyb}(c).
Exposure of a freshly cleaved surface to the ultrahigh vacuum
residual gas for 24 hours  apparently gives rise to $p$-doping of
the surface region: The \be of all surface related features is
reduced by $5-10$ meV.  We can thus more confidently assign the $f$
level with the larger \be to the surface and conclude that the
\fourf level in the pristine sample displays a surface core level
shift of about 10 meV.  With this assignment we can disentangle the
contributions from bulk and surface to the photoemission spectra
and arrive at an estimate for the hybridization strength in both
cases.  The \be of the bulk \flevel around \Gbar provides a lower
limit for the indirect bulk hybridization gap of about 15 meV.  Near
\Xbar we observe that the metallic surface state hybridizes with
an \flevel at  about 13 meV. Interpreting this state as the surface
analogue of the bulk conduction band, we conclude that the surface
hybridization gap amounts to $\sim 9$ meV, see Fig.~\ref{fig:hyb}(d).

This observation completes our picture of the surface electronic
structure of \smhb . We find that hybridization is not only weaker at the
surface, the gap is also shifted away from the Fermi level due to a \scls
of the \fourf level. Consequently, the surface remains metallic at the
lowest temperature.
Similarly, Frantzeskakis \etal  suggested that the
chemical potential does not lie in the hybridization gap and that
therefore the metallic state at \Xbar is not  an in-gap \tss 
but the continuation of the bulk \dband  \cite{Frantzeskakis13}. 
We  can agree with the spirit of their conclusion, but note that
it only applies to the surface region and not to the bulk. 

Regarding the tiny size of the \fourf \scls , we note that for the
B-terminated surface  the chemical environment of the surface Sm
atoms is nominally identical to the bulk. As a consequence, the
\fourf surface shift is naturally much smaller than that of rare
earth metals where the coordination of surface atoms is strongly
reduced compared to their bulk counterparts. In those cases, the
shift amounts to several 100 meV. We observe a shift of that order
of magnitude for the Sm atoms in the outermost layer of Sm-terminated
surfaces (Fig.~\ref{fig:G}(j)).  In this context we also note that
the surface \sHfivetwo multiplet directly confirms the presence of
\smtwopl ions at the surface.  This result clearly conflicts with
the assertion by Phelan \etal , based on X-ray absorption spectroscopy,
that the surface consists of \smthreepl exclusively \cite{Phelan:2014bz}.

Finally, we mention that a change of the $d$-$f$  hybridization
strength at the surface is not unusual and has been observed with
ARPES for other heavy fermion systems \cite{VyalikhPRL09}.  As a
possible cause for weaker surface hybridization in the present
system, we point out that the application of pressure is known to
suppress the bulk gap of   \smhb \cite{Gabani:2003bl}. Experimental
determinations of the surface structure are lacking so far but
calculations indicate an appreciable reduction of the separation
between the outermost layers \cite{Kim:2014hm}.

\section{Discussion} 

\begin{figure}
\centering
\includegraphics[width=.5\linewidth]{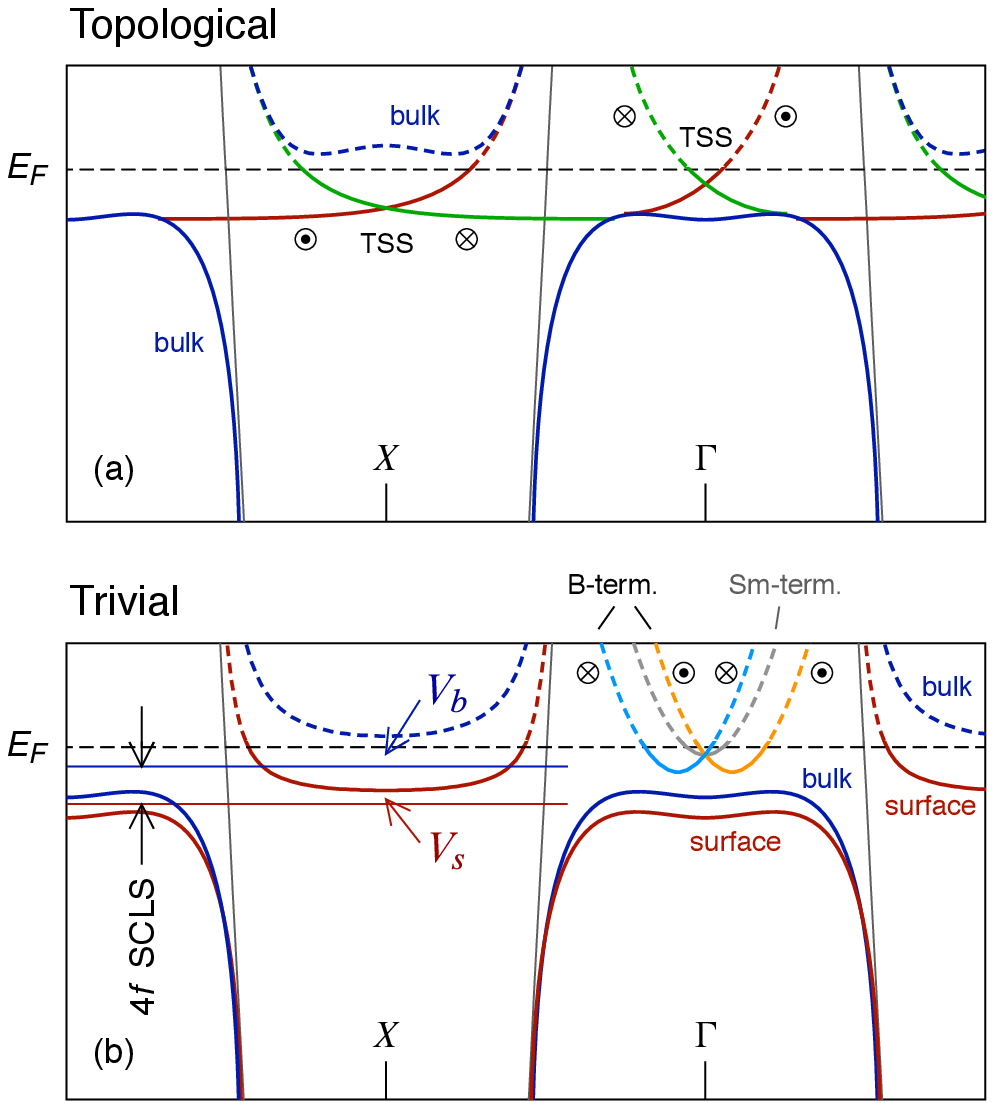}
\caption{(a) Sketch of the situation predicted by theory, and (b)
scenario based on the present experiment.  Instead of the predicted
\tsss (TSS), we observe a massive surface state with a spin degeneracy
that can be lifted by the Rashba effect at \Gbar . Figure (b) further
illustrates how the prominent Fermi surface feature at \Xbar arises
due to a \scls (SCLS) of the \fourf level and a difference in
hybridization strength $V$ between bulk and surface.  Unhybridized
bands are indicated by thin solid lines.}
\label{fig:sketch}
\end{figure}

In summary, we find evidence that strongly supports the existence
of two trivial metallic states that can account for the observed
surface conductivity of \smhb.  In Fig.~\ref{fig:sketch}, we
graphically contrast the theoretically predicted scenario (a) with
the interpretation based on our experimental results (b).
Instead of topological surface states giving rise to Dirac
cones at \Gbar and \Xbar, we find a massive free electron-like state
at \Gbar, and the occupied conduction band of the surface $d$-$f$
hybrid at \Xbar .

While our data are compatible with all previous experimental
reports, there are two rather strong observations apparently
supporting the topological prediction that we will try to counter
in the following. Firstly, Xu \etal have found the helical spin
texture expected for a \tss with spin-resolved ARPES \cite{XuNC14}.
However, we point out that the possibility of a spin-polarized
Rashba-split state of considerable extent in $k$-space, such as we
observe at \Gbar,   has not been considered in the interpretation
of their results. We further note that the energy resolution of
state-of-the-art spin-resolved ARPES is insufficient to distinguish
between the surface state and the \fmult . This issue is exacerbated
by the study's heavy reliance on circularly polarized light, which
is known to give rise to strongly spin-polarized emission from the
\fmult \cite{Suga:2013vu}.  Moreover, it is not clear why spin-resolved
measurements on YbB$_6$ by the same group yield a topological spin
texture for the surface state at \Xbar \cite{XuYbB6}, while a strong
case can be made that the dispersion is that of a trivial state
\cite{FrantzeskakisYbB6}.  The second, although somewhat indirect,
experimental result in support of \tsss in \smhb is provided by
impurity doping experiments.  Kim \etal find that the residual low
temperature conductivity disappears upon doping with magnetic Gd
impurities but not with non-magnetic Y \cite{KimDJSmGdB6}, compatible
with the expectation that a gap should open at the Dirac point of
a \tss upon breaking time reversal symmetry. However, we point out
that such disordered magnetic moments do not open a gap at the Dirac
point of the established topological insulator Bi$_2$Se$_3$ as has
been shown for Fe and Gd \cite{Valla:2012jz,*Scholz:2012kz}. We
therefore conclude that these arguments in favor of the topological
insulator phase in \smhb are insufficiently strong to challenge our
trivial interpretation.

What could explain the chasm between our interpretation and the
theoretical prediction? We iterate that the topological classification
of \smhb hinges on a band inversion at the \Xpoint .  Regardless
of the applied method, the calculated conduction band is found to
have a W-shaped dispersion at $X$ indicative of the inversion
\cite{Takimoto11,LuF13,Alexandrov13,Kim:2014hm}.  Within the limits
of the \kperp resolution achievable in ARPES, we observe a simple
U-shape instead, as shown in Fig.~\ref{fig:hyb}(e--g).

We note with respect to the Rashba splitting of the \Gbar state
that, in spite of the small Rashba parameter \rshbpr, the observed
splitting in momentum space ($\Delta \kpar = 0.08$ \AA$^{-1}$)
easily exceeds that of the largest known semiconductor Rashba systems
\cite{Ishizaka:2011dy,*King:2011bu}.  Heavy-fermion systems could
thus be a fruitful source of materials with very large Rashba
splitting.  Relying on the effective mass instead of the Rashba
parameter to achieve the splitting facilitates control over the
splitting by means of relatively small changes to the surface
potential.

With regard to the surface termination, we note that we observe
apparently pure B and Sm terminations which is surprising since
this would lead to divergence of the electrostatic surface energy
for bulk truncated structures. We cannot provide a comprehensive
picture of the different terminations on the basis of our results
and the current literature.  The umklapp features we observe for
either termination are compatible with the finding by R\" o\ss ler
\etal that most of the surface area exhibits a $(2 \times 1)$
reconstruction \cite{Rossler:2014kn}. However, an interpretation
\cite{Rossler:2014kn} in terms of missing Sm rows is clearly
incompatible with the spectra in Fig.~\ref{fig:G}(i--l).  We note,
though, that both the metallic surface states at \Gbar and \Xbar
are present regardless of the termination. This clearly indicates
that they are not confined to the outermost surface layer, but enjoy
the chemical protection by at least one B layer. We suggest that
this isolation from the environment can account for the observed
robustness \cite{Wolgast:2013ih} of the surface conductivity.

We thank Prof. B. Lake for continuous support of the present activity.
The authors acknowledge financial support by Deutsche
Forschungsgemeinschaft through SPP 1666, by "Impuls- und Vernetzungsfonds
der Helm\-holtz-Gemein\-schaft" through Helm\-holtz-Russia Joint Research
Group no. 408 and Helm\-holtz Virtual Institute "New states of matter
and their excitations".  The authors also acknowledge the support
of Slovak grant agency projects VEGA 2/0106/13, APVV 0132-11, and
CFNT MVEP of the Slovak Academy of Sciences.

\bibliographystyle{aipnum4-1}
\bibliography{full}

\begin{thebibliography}{50}%
\makeatletter
\providecommand \@ifxundefined [1]{%
 \@ifx{#1\undefined}
}%
\providecommand \@ifnum [1]{%
 \ifnum #1\expandafter \@firstoftwo
 \else \expandafter \@secondoftwo
 \fi
}%
\providecommand \@ifx [1]{%
 \ifx #1\expandafter \@firstoftwo
 \else \expandafter \@secondoftwo
 \fi
}%
\providecommand \natexlab [1]{#1}%
\providecommand \enquote  [1]{``#1''}%
\providecommand \bibnamefont  [1]{#1}%
\providecommand \bibfnamefont [1]{#1}%
\providecommand \citenamefont [1]{#1}%
\providecommand \href@noop [0]{\@secondoftwo}%
\providecommand \href [0]{\begingroup \@sanitize@url \@href}%
\providecommand \@href[1]{\@@startlink{#1}\@@href}%
\providecommand \@@href[1]{\endgroup#1\@@endlink}%
\providecommand \@sanitize@url [0]{\catcode `\\12\catcode `\$12\catcode
  `\&12\catcode `\#12\catcode `\^12\catcode `\_12\catcode `\%12\relax}%
\providecommand \@@startlink[1]{}%
\providecommand \@@endlink[0]{}%
\providecommand \url  [0]{\begingroup\@sanitize@url \@url }%
\providecommand \@url [1]{\endgroup\@href {#1}{\urlprefix }}%
\providecommand \urlprefix  [0]{URL }%
\providecommand \Eprint [0]{\href }%
\providecommand \doibase [0]{http://dx.doi.org/}%
\providecommand \selectlanguage [0]{\@gobble}%
\providecommand \bibinfo  [0]{\@secondoftwo}%
\providecommand \bibfield  [0]{\@secondoftwo}%
\providecommand \translation [1]{[#1]}%
\providecommand \BibitemOpen [0]{}%
\providecommand \bibitemStop [0]{}%
\providecommand \bibitemNoStop [0]{.\EOS\space}%
\providecommand \EOS [0]{\spacefactor3000\relax}%
\providecommand \BibitemShut  [1]{\csname bibitem#1\endcsname}%
\let\auto@bib@innerbib\@empty
\bibitem [{\citenamefont {Hasan}\ and\ \citenamefont
  {Kane}(2010)}]{Hasan:2010ku}%
  \BibitemOpen
  \bibfield  {author} {\bibinfo {author} {\bibfnamefont {M.~Z.}\ \bibnamefont
  {Hasan}}\ and\ \bibinfo {author} {\bibfnamefont {C.~L.}\ \bibnamefont
  {Kane}},\ }\href@noop {} {\bibfield  {journal} {\bibinfo  {journal} {Rev.
  Mod. Phys.}\ }\textbf {\bibinfo {volume} {82}},\ \bibinfo {pages} {3045}
  (\bibinfo {year} {2010})}\BibitemShut {NoStop}%
\bibitem [{\citenamefont {Bernevig}\ and\ \citenamefont
  {Hughes}(2013)}]{Bernevig:un}%
  \BibitemOpen
  \bibfield  {author} {\bibinfo {author} {\bibfnamefont {B.~A.}\ \bibnamefont
  {Bernevig}}\ and\ \bibinfo {author} {\bibfnamefont {T.~L.}\ \bibnamefont
  {Hughes}},\ }\href@noop {} {\emph {\bibinfo {title} {{Topological Insulators
  and Topological Superconductors}}}}\ (\bibinfo  {publisher} {Princeton
  University Press},\ \bibinfo {year} {2013})\BibitemShut {NoStop}%
\bibitem [{\citenamefont {Ando}(2013)}]{Ando:2013fg}%
  \BibitemOpen
  \bibfield  {author} {\bibinfo {author} {\bibfnamefont {Y.}~\bibnamefont
  {Ando}},\ }\href@noop {} {\bibfield  {journal} {\bibinfo  {journal} {J. Phys.
  Soc. Jpn.}\ }\textbf {\bibinfo {volume} {82}},\ \bibinfo {pages} {102001}
  (\bibinfo {year} {2013})}\BibitemShut {NoStop}%
\bibitem [{\citenamefont {Shitade}\ \emph {et~al.}(2009)\citenamefont
  {Shitade}, \citenamefont {Katsura}, \citenamefont {Kune{\v s}}, \citenamefont
  {Qi}, \citenamefont {Zhang},\ and\ \citenamefont {Nagaosa}}]{Shitade:2009he}%
  \BibitemOpen
  \bibfield  {author} {\bibinfo {author} {\bibfnamefont {A.}~\bibnamefont
  {Shitade}}, \bibinfo {author} {\bibfnamefont {H.}~\bibnamefont {Katsura}},
  \bibinfo {author} {\bibfnamefont {J.}~\bibnamefont {Kune{\v s}}}, \bibinfo
  {author} {\bibfnamefont {X.-L.}\ \bibnamefont {Qi}}, \bibinfo {author}
  {\bibfnamefont {S.-C.}\ \bibnamefont {Zhang}}, \ and\ \bibinfo {author}
  {\bibfnamefont {N.}~\bibnamefont {Nagaosa}},\ }\href@noop {} {\bibfield
  {journal} {\bibinfo  {journal} {Phys. Rev. Lett.}\ }\textbf {\bibinfo
  {volume} {102}},\ \bibinfo {pages} {256403} (\bibinfo {year}
  {2009})}\BibitemShut {NoStop}%
\bibitem [{\citenamefont {Guo}\ and\ \citenamefont {Franz}(2009)}]{Guo:2009gj}%
  \BibitemOpen
  \bibfield  {author} {\bibinfo {author} {\bibfnamefont {H.-M.}\ \bibnamefont
  {Guo}}\ and\ \bibinfo {author} {\bibfnamefont {M.}~\bibnamefont {Franz}},\
  }\href@noop {} {\bibfield  {journal} {\bibinfo  {journal} {Phys. Rev. Lett.}\
  }\textbf {\bibinfo {volume} {103}},\ \bibinfo {pages} {206805} (\bibinfo
  {year} {2009})}\BibitemShut {NoStop}%
\bibitem [{\citenamefont {Dzero}\ \emph {et~al.}(2010)\citenamefont {Dzero},
  \citenamefont {Sun}, \citenamefont {Galitski},\ and\ \citenamefont
  {Coleman}}]{Dzero10}%
  \BibitemOpen
  \bibfield  {author} {\bibinfo {author} {\bibfnamefont {M.}~\bibnamefont
  {Dzero}}, \bibinfo {author} {\bibfnamefont {K.}~\bibnamefont {Sun}}, \bibinfo
  {author} {\bibfnamefont {V.}~\bibnamefont {Galitski}}, \ and\ \bibinfo
  {author} {\bibfnamefont {P.}~\bibnamefont {Coleman}},\ }\href@noop {}
  {\bibfield  {journal} {\bibinfo  {journal} {Phys. Rev. Lett.}\ }\textbf
  {\bibinfo {volume} {104}},\ \bibinfo {pages} {106408} (\bibinfo {year}
  {2010})}\BibitemShut {NoStop}%
\bibitem [{\citenamefont {Takimoto}(2011)}]{Takimoto11}%
  \BibitemOpen
  \bibfield  {author} {\bibinfo {author} {\bibfnamefont {T.}~\bibnamefont
  {Takimoto}},\ }\href@noop {} {\bibfield  {journal} {\bibinfo  {journal} {J.
  Phys. Soc. Jpn.}\ }\textbf {\bibinfo {volume} {80}},\ \bibinfo {pages}
  {123710} (\bibinfo {year} {2011})}\BibitemShut {NoStop}%
\bibitem [{\citenamefont {Lu}\ \emph {et~al.}(2013)\citenamefont {Lu},
  \citenamefont {Zhao}, \citenamefont {Weng}, \citenamefont {Fang},\ and\
  \citenamefont {Dai}}]{LuF13}%
  \BibitemOpen
  \bibfield  {author} {\bibinfo {author} {\bibfnamefont {F.}~\bibnamefont
  {Lu}}, \bibinfo {author} {\bibfnamefont {J.}~\bibnamefont {Zhao}}, \bibinfo
  {author} {\bibfnamefont {H.}~\bibnamefont {Weng}}, \bibinfo {author}
  {\bibfnamefont {Z.}~\bibnamefont {Fang}}, \ and\ \bibinfo {author}
  {\bibfnamefont {X.}~\bibnamefont {Dai}},\ }\href@noop {} {\bibfield
  {journal} {\bibinfo  {journal} {Phys. Rev. Lett.}\ }\textbf {\bibinfo
  {volume} {110}},\ \bibinfo {pages} {096401} (\bibinfo {year}
  {2013})}\BibitemShut {NoStop}%
\bibitem [{\citenamefont {Alexandrov}, \citenamefont {Dzero},\ and\
  \citenamefont {Coleman}(2013)}]{Alexandrov13}%
  \BibitemOpen
  \bibfield  {author} {\bibinfo {author} {\bibfnamefont {V.}~\bibnamefont
  {Alexandrov}}, \bibinfo {author} {\bibfnamefont {M.}~\bibnamefont {Dzero}}, \
  and\ \bibinfo {author} {\bibfnamefont {P.}~\bibnamefont {Coleman}},\
  }\href@noop {} {\bibfield  {journal} {\bibinfo  {journal} {Phys. Rev. Lett.}\
  }\textbf {\bibinfo {volume} {111}},\ \bibinfo {pages} {226403} (\bibinfo
  {year} {2013})}\BibitemShut {NoStop}%
\bibitem [{\citenamefont {Roy}\ \emph {et~al.}(2014)\citenamefont {Roy},
  \citenamefont {Sau}, \citenamefont {Dzero},\ and\ \citenamefont
  {Galitski}}]{Roy14}%
  \BibitemOpen
  \bibfield  {author} {\bibinfo {author} {\bibfnamefont {B.}~\bibnamefont
  {Roy}}, \bibinfo {author} {\bibfnamefont {J.~D.}\ \bibnamefont {Sau}},
  \bibinfo {author} {\bibfnamefont {M.}~\bibnamefont {Dzero}}, \ and\ \bibinfo
  {author} {\bibfnamefont {V.}~\bibnamefont {Galitski}},\ }\href@noop {}
  {\bibfield  {journal} {\bibinfo  {journal} {Phys. Rev. B}\ }\textbf {\bibinfo
  {volume} {90}},\ \bibinfo {pages} {155314} (\bibinfo {year}
  {2014})}\BibitemShut {NoStop}%
\bibitem [{\citenamefont {Neupane}\ \emph {et~al.}(2013)\citenamefont
  {Neupane}, \citenamefont {Alidoust}, \citenamefont {Xu}, \citenamefont
  {Kondo}, \citenamefont {Ishida}, \citenamefont {Kim}, \citenamefont {Liu},
  \citenamefont {Belopolski}, \citenamefont {Jo}, \citenamefont {Chang},
  \citenamefont {Jeng}, \citenamefont {Durakiewicz}, \citenamefont {Balicas},
  \citenamefont {Lin}, \citenamefont {Bansil}, \citenamefont {Shin},
  \citenamefont {Fisk},\ and\ \citenamefont {Hasan}}]{Neupane13}%
  \BibitemOpen
  \bibfield  {author} {\bibinfo {author} {\bibfnamefont {M.}~\bibnamefont
  {Neupane}}, \bibinfo {author} {\bibfnamefont {N.}~\bibnamefont {Alidoust}},
  \bibinfo {author} {\bibfnamefont {S.-Y.}\ \bibnamefont {Xu}}, \bibinfo
  {author} {\bibfnamefont {T.}~\bibnamefont {Kondo}}, \bibinfo {author}
  {\bibfnamefont {Y.}~\bibnamefont {Ishida}}, \bibinfo {author} {\bibfnamefont
  {D.~J.}\ \bibnamefont {Kim}}, \bibinfo {author} {\bibfnamefont
  {C.}~\bibnamefont {Liu}}, \bibinfo {author} {\bibfnamefont {I.}~\bibnamefont
  {Belopolski}}, \bibinfo {author} {\bibfnamefont {Y.~J.}\ \bibnamefont {Jo}},
  \bibinfo {author} {\bibfnamefont {T.-R.}\ \bibnamefont {Chang}}, \bibinfo
  {author} {\bibfnamefont {H.-T.}\ \bibnamefont {Jeng}}, \bibinfo {author}
  {\bibfnamefont {T.}~\bibnamefont {Durakiewicz}}, \bibinfo {author}
  {\bibfnamefont {L.}~\bibnamefont {Balicas}}, \bibinfo {author} {\bibfnamefont
  {H.}~\bibnamefont {Lin}}, \bibinfo {author} {\bibfnamefont {A.}~\bibnamefont
  {Bansil}}, \bibinfo {author} {\bibfnamefont {S.}~\bibnamefont {Shin}},
  \bibinfo {author} {\bibfnamefont {Z.}~\bibnamefont {Fisk}}, \ and\ \bibinfo
  {author} {\bibfnamefont {M.~Z.}\ \bibnamefont {Hasan}},\ }\href@noop {}
  {\bibfield  {journal} {\bibinfo  {journal} {Nat. Comm.}\ }\textbf {\bibinfo
  {volume} {4}},\ \bibinfo {pages} {2991} (\bibinfo {year} {2013})}\BibitemShut
  {NoStop}%
\bibitem [{\citenamefont {Jiang}\ \emph {et~al.}(2013)\citenamefont {Jiang},
  \citenamefont {Li}, \citenamefont {Zhang}, \citenamefont {Sun}, \citenamefont
  {Chen}, \citenamefont {Ye}, \citenamefont {Xu}, \citenamefont {Ge},
  \citenamefont {Tan}, \citenamefont {Niu}, \citenamefont {Xia}, \citenamefont
  {Xie}, \citenamefont {Li}, \citenamefont {Chen}, \citenamefont {Wen},\ and\
  \citenamefont {Feng}}]{JiangNC13}%
  \BibitemOpen
  \bibfield  {author} {\bibinfo {author} {\bibfnamefont {J.}~\bibnamefont
  {Jiang}}, \bibinfo {author} {\bibfnamefont {S.}~\bibnamefont {Li}}, \bibinfo
  {author} {\bibfnamefont {T.}~\bibnamefont {Zhang}}, \bibinfo {author}
  {\bibfnamefont {Z.}~\bibnamefont {Sun}}, \bibinfo {author} {\bibfnamefont
  {F.}~\bibnamefont {Chen}}, \bibinfo {author} {\bibfnamefont {Z.~R.}\
  \bibnamefont {Ye}}, \bibinfo {author} {\bibfnamefont {M.}~\bibnamefont {Xu}},
  \bibinfo {author} {\bibfnamefont {Q.~Q.}\ \bibnamefont {Ge}}, \bibinfo
  {author} {\bibfnamefont {S.~Y.}\ \bibnamefont {Tan}}, \bibinfo {author}
  {\bibfnamefont {X.~H.}\ \bibnamefont {Niu}}, \bibinfo {author} {\bibfnamefont
  {M.}~\bibnamefont {Xia}}, \bibinfo {author} {\bibfnamefont {B.~P.}\
  \bibnamefont {Xie}}, \bibinfo {author} {\bibfnamefont {Y.~F.}\ \bibnamefont
  {Li}}, \bibinfo {author} {\bibfnamefont {X.~H.}\ \bibnamefont {Chen}},
  \bibinfo {author} {\bibfnamefont {H.~H.}\ \bibnamefont {Wen}}, \ and\
  \bibinfo {author} {\bibfnamefont {D.~L.}\ \bibnamefont {Feng}},\ }\href@noop
  {} {\bibfield  {journal} {\bibinfo  {journal} {Nat. Comm.}\ }\textbf
  {\bibinfo {volume} {4}},\ \bibinfo {pages} {3010} (\bibinfo {year}
  {2013})}\BibitemShut {NoStop}%
\bibitem [{\citenamefont {Xu}\ \emph {et~al.}(2013)\citenamefont {Xu},
  \citenamefont {Shi}, \citenamefont {Biswas}, \citenamefont {Matt},
  \citenamefont {Dhaka}, \citenamefont {Huang}, \citenamefont {Plumb},
  \citenamefont {Radovi{\'c}}, \citenamefont {Dil}, \citenamefont
  {Pomjakushina}, \citenamefont {Conder}, \citenamefont {Amato}, \citenamefont
  {Salman}, \citenamefont {Paul}, \citenamefont {Mesot}, \citenamefont {Ding},\
  and\ \citenamefont {Shi}}]{XuPRB13}%
  \BibitemOpen
  \bibfield  {author} {\bibinfo {author} {\bibfnamefont {N.}~\bibnamefont
  {Xu}}, \bibinfo {author} {\bibfnamefont {X.}~\bibnamefont {Shi}}, \bibinfo
  {author} {\bibfnamefont {P.~K.}\ \bibnamefont {Biswas}}, \bibinfo {author}
  {\bibfnamefont {C.~E.}\ \bibnamefont {Matt}}, \bibinfo {author}
  {\bibfnamefont {R.~S.}\ \bibnamefont {Dhaka}}, \bibinfo {author}
  {\bibfnamefont {Y.}~\bibnamefont {Huang}}, \bibinfo {author} {\bibfnamefont
  {N.~C.}\ \bibnamefont {Plumb}}, \bibinfo {author} {\bibfnamefont
  {M.}~\bibnamefont {Radovi{\'c}}}, \bibinfo {author} {\bibfnamefont {J.~H.}\
  \bibnamefont {Dil}}, \bibinfo {author} {\bibfnamefont {E.}~\bibnamefont
  {Pomjakushina}}, \bibinfo {author} {\bibfnamefont {K.}~\bibnamefont
  {Conder}}, \bibinfo {author} {\bibfnamefont {A.}~\bibnamefont {Amato}},
  \bibinfo {author} {\bibfnamefont {Z.}~\bibnamefont {Salman}}, \bibinfo
  {author} {\bibfnamefont {D.~M.}\ \bibnamefont {Paul}}, \bibinfo {author}
  {\bibfnamefont {J.}~\bibnamefont {Mesot}}, \bibinfo {author} {\bibfnamefont
  {H.}~\bibnamefont {Ding}}, \ and\ \bibinfo {author} {\bibfnamefont
  {M.}~\bibnamefont {Shi}},\ }\href@noop {} {\bibfield  {journal} {\bibinfo
  {journal} {Phys. Rev. B}\ }\textbf {\bibinfo {volume} {88}},\ \bibinfo
  {pages} {121102} (\bibinfo {year} {2013})}\BibitemShut {NoStop}%
\bibitem [{\citenamefont {Xu}\ \emph {et~al.}(2014{\natexlab{a}})\citenamefont
  {Xu}, \citenamefont {Biswas}, \citenamefont {Dil}, \citenamefont {Dhaka},
  \citenamefont {Landolt}, \citenamefont {Muff}, \citenamefont {Matt},
  \citenamefont {Shi}, \citenamefont {Plumb}, \citenamefont {Radovi{\'c}},
  \citenamefont {Pomjakushina}, \citenamefont {Conder}, \citenamefont {Amato},
  \citenamefont {Borisenko}, \citenamefont {Yu}, \citenamefont {Weng},
  \citenamefont {Fang}, \citenamefont {Dai}, \citenamefont {Mesot},
  \citenamefont {Ding},\ and\ \citenamefont {Shi}}]{XuNC14}%
  \BibitemOpen
  \bibfield  {author} {\bibinfo {author} {\bibfnamefont {N.}~\bibnamefont
  {Xu}}, \bibinfo {author} {\bibfnamefont {P.~K.}\ \bibnamefont {Biswas}},
  \bibinfo {author} {\bibfnamefont {J.~H.}\ \bibnamefont {Dil}}, \bibinfo
  {author} {\bibfnamefont {R.~S.}\ \bibnamefont {Dhaka}}, \bibinfo {author}
  {\bibfnamefont {G.}~\bibnamefont {Landolt}}, \bibinfo {author} {\bibfnamefont
  {S.}~\bibnamefont {Muff}}, \bibinfo {author} {\bibfnamefont {C.~E.}\
  \bibnamefont {Matt}}, \bibinfo {author} {\bibfnamefont {X.}~\bibnamefont
  {Shi}}, \bibinfo {author} {\bibfnamefont {N.~C.}\ \bibnamefont {Plumb}},
  \bibinfo {author} {\bibfnamefont {M.}~\bibnamefont {Radovi{\'c}}}, \bibinfo
  {author} {\bibfnamefont {E.}~\bibnamefont {Pomjakushina}}, \bibinfo {author}
  {\bibfnamefont {K.}~\bibnamefont {Conder}}, \bibinfo {author} {\bibfnamefont
  {A.}~\bibnamefont {Amato}}, \bibinfo {author} {\bibfnamefont {S.~V.}\
  \bibnamefont {Borisenko}}, \bibinfo {author} {\bibfnamefont {R.}~\bibnamefont
  {Yu}}, \bibinfo {author} {\bibfnamefont {H.}~\bibnamefont {Weng}}, \bibinfo
  {author} {\bibfnamefont {Z.}~\bibnamefont {Fang}}, \bibinfo {author}
  {\bibfnamefont {X.}~\bibnamefont {Dai}}, \bibinfo {author} {\bibfnamefont
  {J.}~\bibnamefont {Mesot}}, \bibinfo {author} {\bibfnamefont
  {H.}~\bibnamefont {Ding}}, \ and\ \bibinfo {author} {\bibfnamefont
  {M.}~\bibnamefont {Shi}},\ }\href@noop {} {\bibfield  {journal} {\bibinfo
  {journal} {Nat. Comm.}\ }\textbf {\bibinfo {volume} {5}},\ \bibinfo {pages}
  {4566} (\bibinfo {year} {2014}{\natexlab{a}})}\BibitemShut {NoStop}%
\bibitem [{\citenamefont {Min}\ \emph {et~al.}(2014)\citenamefont {Min},
  \citenamefont {Lutz}, \citenamefont {Fiedler}, \citenamefont {Kang},
  \citenamefont {Cho}, \citenamefont {Kim}, \citenamefont {Bentmann},\ and\
  \citenamefont {Reinert}}]{Min14}%
  \BibitemOpen
  \bibfield  {author} {\bibinfo {author} {\bibfnamefont {C.-H.}\ \bibnamefont
  {Min}}, \bibinfo {author} {\bibfnamefont {P.}~\bibnamefont {Lutz}}, \bibinfo
  {author} {\bibfnamefont {S.}~\bibnamefont {Fiedler}}, \bibinfo {author}
  {\bibfnamefont {B.~Y.}\ \bibnamefont {Kang}}, \bibinfo {author}
  {\bibfnamefont {B.~K.}\ \bibnamefont {Cho}}, \bibinfo {author} {\bibfnamefont
  {H.~D.}\ \bibnamefont {Kim}}, \bibinfo {author} {\bibfnamefont
  {H.}~\bibnamefont {Bentmann}}, \ and\ \bibinfo {author} {\bibfnamefont
  {F.}~\bibnamefont {Reinert}},\ }\href@noop {} {\bibfield  {journal} {\bibinfo
   {journal} {Phys. Rev. Lett.}\ }\textbf {\bibinfo {volume} {112}},\ \bibinfo
  {pages} {226402} (\bibinfo {year} {2014})}\BibitemShut {NoStop}%
\bibitem [{\citenamefont {Denlinger}\ \emph
  {et~al.}(2013{\natexlab{a}})\citenamefont {Denlinger}, \citenamefont {Allen},
  \citenamefont {Kang}, \citenamefont {Sun}, \citenamefont {Kim}, \citenamefont
  {Shim}, \citenamefont {Min}, \citenamefont {Kim},\ and\ \citenamefont
  {Fisk}}]{Denlinger13126637}%
  \BibitemOpen
  \bibfield  {author} {\bibinfo {author} {\bibfnamefont {J.~D.}\ \bibnamefont
  {Denlinger}}, \bibinfo {author} {\bibfnamefont {J.~W.}\ \bibnamefont
  {Allen}}, \bibinfo {author} {\bibfnamefont {J.-S.}\ \bibnamefont {Kang}},
  \bibinfo {author} {\bibfnamefont {K.}~\bibnamefont {Sun}}, \bibinfo {author}
  {\bibfnamefont {J.~W.}\ \bibnamefont {Kim}}, \bibinfo {author} {\bibfnamefont
  {J.~H.}\ \bibnamefont {Shim}}, \bibinfo {author} {\bibfnamefont {B.~I.}\
  \bibnamefont {Min}}, \bibinfo {author} {\bibfnamefont {D.-J.}\ \bibnamefont
  {Kim}}, \ and\ \bibinfo {author} {\bibfnamefont {Z.}~\bibnamefont {Fisk}},\
  }\href@noop {} {\bibfield  {journal} {\bibinfo  {journal} {arXiv}\ }
  (\bibinfo {year} {2013}{\natexlab{a}})},\ \Eprint
  {http://arxiv.org/abs/1312.6637v1} {1312.6637v1} \BibitemShut {NoStop}%
\bibitem [{\citenamefont {Li}\ \emph {et~al.}(2014)\citenamefont {Li},
  \citenamefont {Xiang}, \citenamefont {Yu}, \citenamefont {Asaba},
  \citenamefont {Lawson}, \citenamefont {Cai}, \citenamefont {Tinsman},
  \citenamefont {Berkley}, \citenamefont {Wolgast}, \citenamefont {Eo},
  \citenamefont {Kim}, \citenamefont {Kurdak}, \citenamefont {Allen},
  \citenamefont {Sun}, \citenamefont {Chen}, \citenamefont {Wang},
  \citenamefont {Fisk},\ and\ \citenamefont {Li}}]{LiG14}%
  \BibitemOpen
  \bibfield  {author} {\bibinfo {author} {\bibfnamefont {G.}~\bibnamefont
  {Li}}, \bibinfo {author} {\bibfnamefont {Z.}~\bibnamefont {Xiang}}, \bibinfo
  {author} {\bibfnamefont {F.}~\bibnamefont {Yu}}, \bibinfo {author}
  {\bibfnamefont {T.}~\bibnamefont {Asaba}}, \bibinfo {author} {\bibfnamefont
  {B.}~\bibnamefont {Lawson}}, \bibinfo {author} {\bibfnamefont
  {P.}~\bibnamefont {Cai}}, \bibinfo {author} {\bibfnamefont {C.}~\bibnamefont
  {Tinsman}}, \bibinfo {author} {\bibfnamefont {A.}~\bibnamefont {Berkley}},
  \bibinfo {author} {\bibfnamefont {S.}~\bibnamefont {Wolgast}}, \bibinfo
  {author} {\bibfnamefont {Y.~S.}\ \bibnamefont {Eo}}, \bibinfo {author}
  {\bibfnamefont {D.-J.}\ \bibnamefont {Kim}}, \bibinfo {author} {\bibfnamefont
  {C.}~\bibnamefont {Kurdak}}, \bibinfo {author} {\bibfnamefont {J.~W.}\
  \bibnamefont {Allen}}, \bibinfo {author} {\bibfnamefont {K.}~\bibnamefont
  {Sun}}, \bibinfo {author} {\bibfnamefont {X.~H.}\ \bibnamefont {Chen}},
  \bibinfo {author} {\bibfnamefont {Y.~Y.}\ \bibnamefont {Wang}}, \bibinfo
  {author} {\bibfnamefont {Z.}~\bibnamefont {Fisk}}, \ and\ \bibinfo {author}
  {\bibfnamefont {L.}~\bibnamefont {Li}},\ }\href@noop {} {\bibfield  {journal}
  {\bibinfo  {journal} {Science}\ }\textbf {\bibinfo {volume} {346}},\ \bibinfo
  {pages} {1208} (\bibinfo {year} {2014})}\BibitemShut {NoStop}%
\bibitem [{\citenamefont {Cooley}\ \emph {et~al.}(1995)\citenamefont {Cooley},
  \citenamefont {Aronson}, \citenamefont {Fisk},\ and\ \citenamefont
  {Canfield}}]{Cooley95}%
  \BibitemOpen
  \bibfield  {author} {\bibinfo {author} {\bibfnamefont {J.~C.}\ \bibnamefont
  {Cooley}}, \bibinfo {author} {\bibfnamefont {M.~C.}\ \bibnamefont {Aronson}},
  \bibinfo {author} {\bibfnamefont {Z.}~\bibnamefont {Fisk}}, \ and\ \bibinfo
  {author} {\bibfnamefont {P.~C.}\ \bibnamefont {Canfield}},\ }\href@noop {}
  {\bibfield  {journal} {\bibinfo  {journal} {Phys. Rev. Lett.}\ }\textbf
  {\bibinfo {volume} {74}},\ \bibinfo {pages} {1629} (\bibinfo {year}
  {1995})}\BibitemShut {NoStop}%
\bibitem [{\citenamefont {Kikoin}\ and\ \citenamefont
  {Mischenko}(1995)}]{Kikoin95}%
  \BibitemOpen
  \bibfield  {author} {\bibinfo {author} {\bibfnamefont {K.~A.}\ \bibnamefont
  {Kikoin}}\ and\ \bibinfo {author} {\bibfnamefont {A.~S.}\ \bibnamefont
  {Mischenko}},\ }\href@noop {} {\bibfield  {journal} {\bibinfo  {journal} {J
  Phys-Condens Mat}\ }\textbf {\bibinfo {volume} {7}},\ \bibinfo {pages} {307}
  (\bibinfo {year} {1995})}\BibitemShut {NoStop}%
\bibitem [{\citenamefont {Kasuya}(1996)}]{Kasuya96}%
  \BibitemOpen
  \bibfield  {author} {\bibinfo {author} {\bibfnamefont {T.}~\bibnamefont
  {Kasuya}},\ }\href@noop {} {\bibfield  {journal} {\bibinfo  {journal} {J.
  Phys. Soc. Jpn.}\ }\textbf {\bibinfo {volume} {65}},\ \bibinfo {pages} {2548}
  (\bibinfo {year} {1996})}\BibitemShut {NoStop}%
\bibitem [{\citenamefont {Riseborough}(2000)}]{Riseborough00}%
  \BibitemOpen
  \bibfield  {author} {\bibinfo {author} {\bibfnamefont {P.~S.}\ \bibnamefont
  {Riseborough}},\ }\href@noop {} {\bibfield  {journal} {\bibinfo  {journal}
  {Adv Phys}\ }\textbf {\bibinfo {volume} {49}},\ \bibinfo {pages} {257}
  (\bibinfo {year} {2000})}\BibitemShut {NoStop}%
\bibitem [{\citenamefont {Sluchanko}\ \emph {et~al.}(2000)\citenamefont
  {Sluchanko}, \citenamefont {Glushkov}, \citenamefont {Gorshunov},
  \citenamefont {Demishev}, \citenamefont {Kondrin}, \citenamefont {Pronin},
  \citenamefont {Volkov}, \citenamefont {Savchenko}, \citenamefont
  {Gr{\"u}ner}, \citenamefont {Bruynseraede}, \citenamefont {Moshchalkov},\
  and\ \citenamefont {Kunii}}]{Sluchanko00}%
  \BibitemOpen
  \bibfield  {author} {\bibinfo {author} {\bibfnamefont {N.~E.}\ \bibnamefont
  {Sluchanko}}, \bibinfo {author} {\bibfnamefont {V.~V.}\ \bibnamefont
  {Glushkov}}, \bibinfo {author} {\bibfnamefont {B.~P.}\ \bibnamefont
  {Gorshunov}}, \bibinfo {author} {\bibfnamefont {S.~V.}\ \bibnamefont
  {Demishev}}, \bibinfo {author} {\bibfnamefont {M.~V.}\ \bibnamefont
  {Kondrin}}, \bibinfo {author} {\bibfnamefont {A.~A.}\ \bibnamefont {Pronin}},
  \bibinfo {author} {\bibfnamefont {A.~A.}\ \bibnamefont {Volkov}}, \bibinfo
  {author} {\bibfnamefont {A.~K.}\ \bibnamefont {Savchenko}}, \bibinfo {author}
  {\bibfnamefont {G.}~\bibnamefont {Gr{\"u}ner}}, \bibinfo {author}
  {\bibfnamefont {Y.}~\bibnamefont {Bruynseraede}}, \bibinfo {author}
  {\bibfnamefont {V.~V.}\ \bibnamefont {Moshchalkov}}, \ and\ \bibinfo {author}
  {\bibfnamefont {S.}~\bibnamefont {Kunii}},\ }\href@noop {} {\bibfield
  {journal} {\bibinfo  {journal} {Phys. Rev. B}\ }\textbf {\bibinfo {volume}
  {61}},\ \bibinfo {pages} {9906} (\bibinfo {year} {2000})}\BibitemShut
  {NoStop}%
\bibitem [{\citenamefont {Vainshtein}, \citenamefont {Blokhin},\ and\
  \citenamefont {Paderno}(1965)}]{Vainshtein65}%
  \BibitemOpen
  \bibfield  {author} {\bibinfo {author} {\bibfnamefont {E.~E.}\ \bibnamefont
  {Vainshtein}}, \bibinfo {author} {\bibfnamefont {S.~M.}\ \bibnamefont
  {Blokhin}}, \ and\ \bibinfo {author} {\bibfnamefont {Y.~B.}\ \bibnamefont
  {Paderno}},\ }\href@noop {} {\bibfield  {journal} {\bibinfo  {journal} {Sov.
  Phys. Solid State}\ }\textbf {\bibinfo {volume} {6}},\ \bibinfo {pages}
  {2318} (\bibinfo {year} {1965})}\BibitemShut {NoStop}%
\bibitem [{\citenamefont {Menth}, \citenamefont {Buehler},\ and\ \citenamefont
  {Geballe}(1969)}]{Menth69}%
  \BibitemOpen
  \bibfield  {author} {\bibinfo {author} {\bibfnamefont {A.}~\bibnamefont
  {Menth}}, \bibinfo {author} {\bibfnamefont {E.}~\bibnamefont {Buehler}}, \
  and\ \bibinfo {author} {\bibfnamefont {T.~H.}\ \bibnamefont {Geballe}},\
  }\href@noop {} {\bibfield  {journal} {\bibinfo  {journal} {Phys. Rev. Lett.}\
  }\textbf {\bibinfo {volume} {22}},\ \bibinfo {pages} {295} (\bibinfo {year}
  {1969})}\BibitemShut {NoStop}%
\bibitem [{\citenamefont {Nickerson}\ \emph {et~al.}(1971)\citenamefont
  {Nickerson}, \citenamefont {White}, \citenamefont {Lee}, \citenamefont
  {Bachmann}, \citenamefont {Geballe},\ and\ \citenamefont
  {Hull}}]{Nickerson71}%
  \BibitemOpen
  \bibfield  {author} {\bibinfo {author} {\bibfnamefont {J.~C.}\ \bibnamefont
  {Nickerson}}, \bibinfo {author} {\bibfnamefont {R.~M.}\ \bibnamefont
  {White}}, \bibinfo {author} {\bibfnamefont {K.~N.}\ \bibnamefont {Lee}},
  \bibinfo {author} {\bibfnamefont {R.}~\bibnamefont {Bachmann}}, \bibinfo
  {author} {\bibfnamefont {T.~H.}\ \bibnamefont {Geballe}}, \ and\ \bibinfo
  {author} {\bibfnamefont {G.~W.}\ \bibnamefont {Hull}, \bibfnamefont {Jr}},\
  }\href@noop {} {\bibfield  {journal} {\bibinfo  {journal} {Phys. Rev. B}\
  }\textbf {\bibinfo {volume} {3}},\ \bibinfo {pages} {2030} (\bibinfo {year}
  {1971})}\BibitemShut {NoStop}%
\bibitem [{\citenamefont {Allen}, \citenamefont {Batlogg},\ and\ \citenamefont
  {Wachter}(1979)}]{Allen79}%
  \BibitemOpen
  \bibfield  {author} {\bibinfo {author} {\bibfnamefont {J.~W.}\ \bibnamefont
  {Allen}}, \bibinfo {author} {\bibfnamefont {B.}~\bibnamefont {Batlogg}}, \
  and\ \bibinfo {author} {\bibfnamefont {P.}~\bibnamefont {Wachter}},\
  }\href@noop {} {\bibfield  {journal} {\bibinfo  {journal} {Phys. Rev. B}\
  }\textbf {\bibinfo {volume} {20}},\ \bibinfo {pages} {4807} (\bibinfo {year}
  {1979})}\BibitemShut {NoStop}%
\bibitem [{\citenamefont {Wolgast}\ \emph {et~al.}(2013)\citenamefont
  {Wolgast}, \citenamefont {Kurdak}, \citenamefont {Sun}, \citenamefont
  {Allen}, \citenamefont {Kim},\ and\ \citenamefont {Fisk}}]{Wolgast:2013ih}%
  \BibitemOpen
  \bibfield  {author} {\bibinfo {author} {\bibfnamefont {S.}~\bibnamefont
  {Wolgast}}, \bibinfo {author} {\bibfnamefont {{\c C}.}~\bibnamefont
  {Kurdak}}, \bibinfo {author} {\bibfnamefont {K.}~\bibnamefont {Sun}},
  \bibinfo {author} {\bibfnamefont {J.~W.}\ \bibnamefont {Allen}}, \bibinfo
  {author} {\bibfnamefont {D.-J.}\ \bibnamefont {Kim}}, \ and\ \bibinfo
  {author} {\bibfnamefont {Z.}~\bibnamefont {Fisk}},\ }\href@noop {} {\bibfield
   {journal} {\bibinfo  {journal} {Phys. Rev. B}\ }\textbf {\bibinfo {volume}
  {88}},\ \bibinfo {pages} {180405} (\bibinfo {year} {2013})}\BibitemShut
  {NoStop}%
\bibitem [{\citenamefont {Zhang}\ \emph {et~al.}(2013)\citenamefont {Zhang},
  \citenamefont {Butch}, \citenamefont {Syers}, \citenamefont {Ziemak},
  \citenamefont {Greene},\ and\ \citenamefont {Paglione}}]{ZhangXPRX13}%
  \BibitemOpen
  \bibfield  {author} {\bibinfo {author} {\bibfnamefont {X.}~\bibnamefont
  {Zhang}}, \bibinfo {author} {\bibfnamefont {N.~P.}\ \bibnamefont {Butch}},
  \bibinfo {author} {\bibfnamefont {P.}~\bibnamefont {Syers}}, \bibinfo
  {author} {\bibfnamefont {S.}~\bibnamefont {Ziemak}}, \bibinfo {author}
  {\bibfnamefont {R.~L.}\ \bibnamefont {Greene}}, \ and\ \bibinfo {author}
  {\bibfnamefont {J.}~\bibnamefont {Paglione}},\ }\href@noop {} {\bibfield
  {journal} {\bibinfo  {journal} {Phys. Rev. X}\ }\textbf {\bibinfo {volume}
  {3}},\ \bibinfo {pages} {011011} (\bibinfo {year} {2013})}\BibitemShut
  {NoStop}%
\bibitem [{\citenamefont {Kim}, \citenamefont {Xia},\ and\ \citenamefont
  {Fisk}(2014)}]{KimDJSmGdB6}%
  \BibitemOpen
  \bibfield  {author} {\bibinfo {author} {\bibfnamefont {D.~J.}\ \bibnamefont
  {Kim}}, \bibinfo {author} {\bibfnamefont {J.}~\bibnamefont {Xia}}, \ and\
  \bibinfo {author} {\bibfnamefont {Z.}~\bibnamefont {Fisk}},\ }\href@noop {}
  {\bibfield  {journal} {\bibinfo  {journal} {Nat. Mater.}\ }\textbf {\bibinfo
  {volume} {13}},\ \bibinfo {pages} {466} (\bibinfo {year} {2014})}\BibitemShut
  {NoStop}%
\bibitem [{\citenamefont {Frantzeskakis}\ \emph {et~al.}(2013)\citenamefont
  {Frantzeskakis}, \citenamefont {de~Jong}, \citenamefont {Zwartsenberg},
  \citenamefont {Huang}, \citenamefont {Pan}, \citenamefont {Zhang},
  \citenamefont {Zhang}, \citenamefont {Zhang}, \citenamefont {Bao},
  \citenamefont {Tegus}, \citenamefont {Varykhalov}, \citenamefont
  {de~Visser},\ and\ \citenamefont {Golden}}]{Frantzeskakis13}%
  \BibitemOpen
  \bibfield  {author} {\bibinfo {author} {\bibfnamefont {E.}~\bibnamefont
  {Frantzeskakis}}, \bibinfo {author} {\bibfnamefont {N.}~\bibnamefont
  {de~Jong}}, \bibinfo {author} {\bibfnamefont {B.}~\bibnamefont
  {Zwartsenberg}}, \bibinfo {author} {\bibfnamefont {Y.~K.}\ \bibnamefont
  {Huang}}, \bibinfo {author} {\bibfnamefont {Y.}~\bibnamefont {Pan}}, \bibinfo
  {author} {\bibfnamefont {X.}~\bibnamefont {Zhang}}, \bibinfo {author}
  {\bibfnamefont {J.~X.}\ \bibnamefont {Zhang}}, \bibinfo {author}
  {\bibfnamefont {F.~X.}\ \bibnamefont {Zhang}}, \bibinfo {author}
  {\bibfnamefont {L.~H.}\ \bibnamefont {Bao}}, \bibinfo {author} {\bibfnamefont
  {O.}~\bibnamefont {Tegus}}, \bibinfo {author} {\bibfnamefont
  {A.}~\bibnamefont {Varykhalov}}, \bibinfo {author} {\bibfnamefont
  {A.}~\bibnamefont {de~Visser}}, \ and\ \bibinfo {author} {\bibfnamefont
  {M.~S.}\ \bibnamefont {Golden}},\ }\href@noop {} {\bibfield  {journal}
  {\bibinfo  {journal} {Phys. Rev. X}\ }\textbf {\bibinfo {volume} {3}},\
  \bibinfo {pages} {041024} (\bibinfo {year} {2013})}\BibitemShut {NoStop}%
\bibitem [{Note1()}]{Note1}%
  \BibitemOpen
  \bibinfo {note} {We find that samples cleave with two distinct terminations
  as is further described in the Methods section.}\BibitemShut {Stop}%
\bibitem [{\citenamefont {Gerken}(1983)}]{Gerken:1983wk}%
  \BibitemOpen
  \bibfield  {author} {\bibinfo {author} {\bibfnamefont {F.}~\bibnamefont
  {Gerken}},\ }\href@noop {} {\bibfield  {journal} {\bibinfo  {journal} {J.
  Phys. F}\ }\textbf {\bibinfo {volume} {13}},\ \bibinfo {pages} {703}
  (\bibinfo {year} {1983})}\BibitemShut {NoStop}%
\bibitem [{Note2()}]{Note2}%
  \BibitemOpen
  \bibinfo {note} {See e.g.,\protect \xspace \protect \href
  {http://iopscience.iop.org/1367-2630/focus/Focus\%20on\%20the\%20Rashba\%20Effect/}{New
  J. Phys., Focus on the Rashba Effect.}}\BibitemShut {Stop}%
\bibitem [{\citenamefont {Chen}\ \emph {et~al.}(2009)\citenamefont {Chen},
  \citenamefont {Analytis}, \citenamefont {Chu}, \citenamefont {Liu},
  \citenamefont {Mo}, \citenamefont {Qi}, \citenamefont {Zhang}, \citenamefont
  {Lu}, \citenamefont {Dai}, \citenamefont {Fang}, \citenamefont {Zhang},
  \citenamefont {Fisher}, \citenamefont {Hussain},\ and\ \citenamefont
  {Shen}}]{Chen:2009do}%
  \BibitemOpen
  \bibfield  {author} {\bibinfo {author} {\bibfnamefont {Y.~L.}\ \bibnamefont
  {Chen}}, \bibinfo {author} {\bibfnamefont {J.~G.}\ \bibnamefont {Analytis}},
  \bibinfo {author} {\bibfnamefont {J.~H.}\ \bibnamefont {Chu}}, \bibinfo
  {author} {\bibfnamefont {Z.~K.}\ \bibnamefont {Liu}}, \bibinfo {author}
  {\bibfnamefont {S.~K.}\ \bibnamefont {Mo}}, \bibinfo {author} {\bibfnamefont
  {X.~L.}\ \bibnamefont {Qi}}, \bibinfo {author} {\bibfnamefont {H.~J.}\
  \bibnamefont {Zhang}}, \bibinfo {author} {\bibfnamefont {D.~H.}\ \bibnamefont
  {Lu}}, \bibinfo {author} {\bibfnamefont {X.}~\bibnamefont {Dai}}, \bibinfo
  {author} {\bibfnamefont {Z.}~\bibnamefont {Fang}}, \bibinfo {author}
  {\bibfnamefont {S.~C.}\ \bibnamefont {Zhang}}, \bibinfo {author}
  {\bibfnamefont {I.~R.}\ \bibnamefont {Fisher}}, \bibinfo {author}
  {\bibfnamefont {Z.}~\bibnamefont {Hussain}}, \ and\ \bibinfo {author}
  {\bibfnamefont {Z.-X.}\ \bibnamefont {Shen}},\ }\href@noop {} {\bibfield
  {journal} {\bibinfo  {journal} {Science}\ }\textbf {\bibinfo {volume}
  {325}},\ \bibinfo {pages} {178} (\bibinfo {year} {2009})}\BibitemShut
  {NoStop}%
\bibitem [{\citenamefont {Nicolay}\ \emph {et~al.}(2001)\citenamefont
  {Nicolay}, \citenamefont {Reinert}, \citenamefont {H{\"u}fner},\ and\
  \citenamefont {Blaha}}]{Nicolay:2001fy}%
  \BibitemOpen
  \bibfield  {author} {\bibinfo {author} {\bibfnamefont {G.}~\bibnamefont
  {Nicolay}}, \bibinfo {author} {\bibfnamefont {F.}~\bibnamefont {Reinert}},
  \bibinfo {author} {\bibfnamefont {S.}~\bibnamefont {H{\"u}fner}}, \ and\
  \bibinfo {author} {\bibfnamefont {P.}~\bibnamefont {Blaha}},\ }\href@noop {}
  {\bibfield  {journal} {\bibinfo  {journal} {Phys. Rev. B}\ }\textbf {\bibinfo
  {volume} {65}},\ \bibinfo {pages} {033407} (\bibinfo {year}
  {2001})}\BibitemShut {NoStop}%
\bibitem [{\citenamefont {Bihlmayer}\ \emph {et~al.}(2006)\citenamefont
  {Bihlmayer}, \citenamefont {Koroteev}, \citenamefont {Echenique},
  \citenamefont {Chulkov},\ and\ \citenamefont {Bl{\"u}gel}}]{Bihlmayer06}%
  \BibitemOpen
  \bibfield  {author} {\bibinfo {author} {\bibfnamefont {G.}~\bibnamefont
  {Bihlmayer}}, \bibinfo {author} {\bibfnamefont {Y.~M.}\ \bibnamefont
  {Koroteev}}, \bibinfo {author} {\bibfnamefont {P.~M.}\ \bibnamefont
  {Echenique}}, \bibinfo {author} {\bibfnamefont {E.~V.}\ \bibnamefont
  {Chulkov}}, \ and\ \bibinfo {author} {\bibfnamefont {S.}~\bibnamefont
  {Bl{\"u}gel}},\ }\href@noop {} {\bibfield  {journal} {\bibinfo  {journal}
  {Surface Science}\ }\textbf {\bibinfo {volume} {600}},\ \bibinfo {pages}
  {3888} (\bibinfo {year} {2006})}\BibitemShut {NoStop}%
\bibitem [{\citenamefont {Varykhalov}\ \emph {et~al.}(2012)\citenamefont
  {Varykhalov}, \citenamefont {Marchenko}, \citenamefont {Scholz},
  \citenamefont {Rienks}, \citenamefont {Kim}, \citenamefont {Bihlmayer},
  \citenamefont {S{\'a}nchez-Barriga},\ and\ \citenamefont
  {Rader}}]{Varykhalov:2012ec}%
  \BibitemOpen
  \bibfield  {author} {\bibinfo {author} {\bibfnamefont {A.}~\bibnamefont
  {Varykhalov}}, \bibinfo {author} {\bibfnamefont {D.}~\bibnamefont
  {Marchenko}}, \bibinfo {author} {\bibfnamefont {M.~R.}\ \bibnamefont
  {Scholz}}, \bibinfo {author} {\bibfnamefont {E.~D.~L.}\ \bibnamefont
  {Rienks}}, \bibinfo {author} {\bibfnamefont {T.~K.}\ \bibnamefont {Kim}},
  \bibinfo {author} {\bibfnamefont {G.}~\bibnamefont {Bihlmayer}}, \bibinfo
  {author} {\bibfnamefont {J.}~\bibnamefont {S{\'a}nchez-Barriga}}, \ and\
  \bibinfo {author} {\bibfnamefont {O.}~\bibnamefont {Rader}},\ }\href@noop {}
  {\bibfield  {journal} {\bibinfo  {journal} {Phys. Rev. Lett.}\ }\textbf
  {\bibinfo {volume} {108}},\ \bibinfo {pages} {066804} (\bibinfo {year}
  {2012})}\BibitemShut {NoStop}%
\bibitem [{\citenamefont {Phelan}\ \emph {et~al.}(2014)\citenamefont {Phelan},
  \citenamefont {Koohpayeh}, \citenamefont {Cottingham}, \citenamefont
  {Freeland}, \citenamefont {Leiner}, \citenamefont {Broholm},\ and\
  \citenamefont {McQueen}}]{Phelan:2014bz}%
  \BibitemOpen
  \bibfield  {author} {\bibinfo {author} {\bibfnamefont {W.~A.}\ \bibnamefont
  {Phelan}}, \bibinfo {author} {\bibfnamefont {S.~M.}\ \bibnamefont
  {Koohpayeh}}, \bibinfo {author} {\bibfnamefont {P.}~\bibnamefont
  {Cottingham}}, \bibinfo {author} {\bibfnamefont {J.~W.}\ \bibnamefont
  {Freeland}}, \bibinfo {author} {\bibfnamefont {J.~C.}\ \bibnamefont
  {Leiner}}, \bibinfo {author} {\bibfnamefont {C.~L.}\ \bibnamefont {Broholm}},
  \ and\ \bibinfo {author} {\bibfnamefont {T.~M.}\ \bibnamefont {McQueen}},\
  }\href@noop {} {\bibfield  {journal} {\bibinfo  {journal} {Phys. Rev. X}\
  }\textbf {\bibinfo {volume} {4}},\ \bibinfo {pages} {031012} (\bibinfo {year}
  {2014})}\BibitemShut {NoStop}%
\bibitem [{\citenamefont {Vyalikh}\ \emph {et~al.}(2009)\citenamefont
  {Vyalikh}, \citenamefont {Danzenb{\"a}cher}, \citenamefont {Kucherenko},
  \citenamefont {Krellner}, \citenamefont {Geibel}, \citenamefont {Laubschat},
  \citenamefont {Shi}, \citenamefont {Patthey}, \citenamefont {Follath},\ and\
  \citenamefont {Molodtsov}}]{VyalikhPRL09}%
  \BibitemOpen
  \bibfield  {author} {\bibinfo {author} {\bibfnamefont {D.~V.}\ \bibnamefont
  {Vyalikh}}, \bibinfo {author} {\bibfnamefont {S.}~\bibnamefont
  {Danzenb{\"a}cher}}, \bibinfo {author} {\bibfnamefont {Y.}~\bibnamefont
  {Kucherenko}}, \bibinfo {author} {\bibfnamefont {C.}~\bibnamefont
  {Krellner}}, \bibinfo {author} {\bibfnamefont {C.}~\bibnamefont {Geibel}},
  \bibinfo {author} {\bibfnamefont {C.}~\bibnamefont {Laubschat}}, \bibinfo
  {author} {\bibfnamefont {M.}~\bibnamefont {Shi}}, \bibinfo {author}
  {\bibfnamefont {L.}~\bibnamefont {Patthey}}, \bibinfo {author} {\bibfnamefont
  {R.}~\bibnamefont {Follath}}, \ and\ \bibinfo {author} {\bibfnamefont
  {S.}~\bibnamefont {Molodtsov}},\ }\href@noop {} {\bibfield  {journal}
  {\bibinfo  {journal} {Phys. Rev. Lett.}\ }\textbf {\bibinfo {volume} {103}},\
  \bibinfo {pages} {137601} (\bibinfo {year} {2009})}\BibitemShut {NoStop}%
\bibitem [{\citenamefont {Gab{\'a}ni}\ \emph {et~al.}(2003)\citenamefont
  {Gab{\'a}ni}, \citenamefont {Bauer}, \citenamefont {Berger}, \citenamefont
  {Flachbart}, \citenamefont {Paderno}, \citenamefont {Paul}, \citenamefont
  {Pavl{\'\i}k},\ and\ \citenamefont {Shitsevalova}}]{Gabani:2003bl}%
  \BibitemOpen
  \bibfield  {author} {\bibinfo {author} {\bibfnamefont {S.}~\bibnamefont
  {Gab{\'a}ni}}, \bibinfo {author} {\bibfnamefont {E.}~\bibnamefont {Bauer}},
  \bibinfo {author} {\bibfnamefont {S.}~\bibnamefont {Berger}}, \bibinfo
  {author} {\bibfnamefont {K.}~\bibnamefont {Flachbart}}, \bibinfo {author}
  {\bibfnamefont {Y.}~\bibnamefont {Paderno}}, \bibinfo {author} {\bibfnamefont
  {C.}~\bibnamefont {Paul}}, \bibinfo {author} {\bibfnamefont {V.}~\bibnamefont
  {Pavl{\'\i}k}}, \ and\ \bibinfo {author} {\bibfnamefont {N.}~\bibnamefont
  {Shitsevalova}},\ }\href@noop {} {\bibfield  {journal} {\bibinfo  {journal}
  {Phys. Rev. B}\ }\textbf {\bibinfo {volume} {67}},\ \bibinfo {pages} {172406}
  (\bibinfo {year} {2003})}\BibitemShut {NoStop}%
\bibitem [{\citenamefont {Kim}\ \emph {et~al.}(2014)\citenamefont {Kim},
  \citenamefont {Kim}, \citenamefont {Kang}, \citenamefont {Kim}, \citenamefont
  {Choi}, \citenamefont {Kang}, \citenamefont {Denlinger},\ and\ \citenamefont
  {Min}}]{Kim:2014hm}%
  \BibitemOpen
  \bibfield  {author} {\bibinfo {author} {\bibfnamefont {J.}~\bibnamefont
  {Kim}}, \bibinfo {author} {\bibfnamefont {K.}~\bibnamefont {Kim}}, \bibinfo
  {author} {\bibfnamefont {C.-J.}\ \bibnamefont {Kang}}, \bibinfo {author}
  {\bibfnamefont {S.}~\bibnamefont {Kim}}, \bibinfo {author} {\bibfnamefont
  {H.~C.}\ \bibnamefont {Choi}}, \bibinfo {author} {\bibfnamefont {J.-S.}\
  \bibnamefont {Kang}}, \bibinfo {author} {\bibfnamefont {J.~D.}\ \bibnamefont
  {Denlinger}}, \ and\ \bibinfo {author} {\bibfnamefont {B.~I.}\ \bibnamefont
  {Min}},\ }\href@noop {} {\bibfield  {journal} {\bibinfo  {journal} {Phys.
  Rev. B}\ }\textbf {\bibinfo {volume} {90}},\ \bibinfo {pages} {075131}
  (\bibinfo {year} {2014})}\BibitemShut {NoStop}%
\bibitem [{\citenamefont {Suga}\ \emph {et~al.}(2013)\citenamefont {Suga},
  \citenamefont {Sakamoto}, \citenamefont {Okuda}, \citenamefont {Miyamoto},
  \citenamefont {Kuroda}, \citenamefont {Sekiyama}, \citenamefont {Yamaguchi},
  \citenamefont {Fujiwara}, \citenamefont {Irizawa},\ and\ \citenamefont
  {Ito}}]{Suga:2013vu}%
  \BibitemOpen
  \bibfield  {author} {\bibinfo {author} {\bibfnamefont {S.}~\bibnamefont
  {Suga}}, \bibinfo {author} {\bibfnamefont {K.}~\bibnamefont {Sakamoto}},
  \bibinfo {author} {\bibfnamefont {T.}~\bibnamefont {Okuda}}, \bibinfo
  {author} {\bibfnamefont {K.}~\bibnamefont {Miyamoto}}, \bibinfo {author}
  {\bibfnamefont {K.}~\bibnamefont {Kuroda}}, \bibinfo {author} {\bibfnamefont
  {A.}~\bibnamefont {Sekiyama}}, \bibinfo {author} {\bibfnamefont
  {J.}~\bibnamefont {Yamaguchi}}, \bibinfo {author} {\bibfnamefont
  {H.}~\bibnamefont {Fujiwara}}, \bibinfo {author} {\bibfnamefont
  {A.}~\bibnamefont {Irizawa}}, \ and\ \bibinfo {author} {\bibfnamefont
  {T.}~\bibnamefont {Ito}},\ }\href@noop {} {\bibfield  {journal} {\bibinfo
  {journal} {J. Phys. Soc. Jpn.}\ }\textbf {\bibinfo {volume} {83}} (\bibinfo
  {year} {2013})}\BibitemShut {NoStop}%
\bibitem [{\citenamefont {Xu}\ \emph {et~al.}(2014{\natexlab{b}})\citenamefont
  {Xu}, \citenamefont {Matt}, \citenamefont {Pomjakushina}, \citenamefont
  {Dil}, \citenamefont {Landolt}, \citenamefont {Ma}, \citenamefont {Shi},
  \citenamefont {Dhaka}, \citenamefont {Plumb}, \citenamefont {Radovi{\'c}},
  \citenamefont {Strocov}, \citenamefont {Kim}, \citenamefont {Hoesch},
  \citenamefont {Conder}, \citenamefont {Mesot}, \citenamefont {Ding},\ and\
  \citenamefont {Shi}}]{XuYbB6}%
  \BibitemOpen
  \bibfield  {author} {\bibinfo {author} {\bibfnamefont {N.}~\bibnamefont
  {Xu}}, \bibinfo {author} {\bibfnamefont {C.~E.}\ \bibnamefont {Matt}},
  \bibinfo {author} {\bibfnamefont {E.}~\bibnamefont {Pomjakushina}}, \bibinfo
  {author} {\bibfnamefont {J.~H.}\ \bibnamefont {Dil}}, \bibinfo {author}
  {\bibfnamefont {G.}~\bibnamefont {Landolt}}, \bibinfo {author} {\bibfnamefont
  {J.~Z.}\ \bibnamefont {Ma}}, \bibinfo {author} {\bibfnamefont
  {X.}~\bibnamefont {Shi}}, \bibinfo {author} {\bibfnamefont {R.~S.}\
  \bibnamefont {Dhaka}}, \bibinfo {author} {\bibfnamefont {N.~C.}\ \bibnamefont
  {Plumb}}, \bibinfo {author} {\bibfnamefont {M.}~\bibnamefont {Radovi{\'c}}},
  \bibinfo {author} {\bibfnamefont {V.~N.}\ \bibnamefont {Strocov}}, \bibinfo
  {author} {\bibfnamefont {T.~K.}\ \bibnamefont {Kim}}, \bibinfo {author}
  {\bibfnamefont {M.}~\bibnamefont {Hoesch}}, \bibinfo {author} {\bibfnamefont
  {K.}~\bibnamefont {Conder}}, \bibinfo {author} {\bibfnamefont
  {J.}~\bibnamefont {Mesot}}, \bibinfo {author} {\bibfnamefont
  {H.}~\bibnamefont {Ding}}, \ and\ \bibinfo {author} {\bibfnamefont
  {M.}~\bibnamefont {Shi}},\ }\href@noop {} {\bibfield  {journal} {\bibinfo
  {journal} {arXiv}\ } (\bibinfo {year} {2014}{\natexlab{b}})},\ \Eprint
  {http://arxiv.org/abs/1405.0165v1} {1405.0165v1} \BibitemShut {NoStop}%
\bibitem [{\citenamefont {Frantzeskakis}\ \emph {et~al.}(2014)\citenamefont
  {Frantzeskakis}, \citenamefont {de~Jong}, \citenamefont {Zhang},
  \citenamefont {Zhang}, \citenamefont {Li}, \citenamefont {Liang},
  \citenamefont {Wang}, \citenamefont {Varykhalov}, \citenamefont {Huang},\
  and\ \citenamefont {Golden}}]{FrantzeskakisYbB6}%
  \BibitemOpen
  \bibfield  {author} {\bibinfo {author} {\bibfnamefont {E.}~\bibnamefont
  {Frantzeskakis}}, \bibinfo {author} {\bibfnamefont {N.}~\bibnamefont
  {de~Jong}}, \bibinfo {author} {\bibfnamefont {J.~X.}\ \bibnamefont {Zhang}},
  \bibinfo {author} {\bibfnamefont {X.}~\bibnamefont {Zhang}}, \bibinfo
  {author} {\bibfnamefont {Z.}~\bibnamefont {Li}}, \bibinfo {author}
  {\bibfnamefont {C.~L.}\ \bibnamefont {Liang}}, \bibinfo {author}
  {\bibfnamefont {Y.}~\bibnamefont {Wang}}, \bibinfo {author} {\bibfnamefont
  {A.}~\bibnamefont {Varykhalov}}, \bibinfo {author} {\bibfnamefont {Y.~K.}\
  \bibnamefont {Huang}}, \ and\ \bibinfo {author} {\bibfnamefont {M.~S.}\
  \bibnamefont {Golden}},\ }\href@noop {} {\bibfield  {journal} {\bibinfo
  {journal} {Phys. Rev. B}\ }\textbf {\bibinfo {volume} {90}},\ \bibinfo
  {pages} {235116} (\bibinfo {year} {2014})}\BibitemShut {NoStop}%
\bibitem [{\citenamefont {Valla}\ \emph {et~al.}(2012)\citenamefont {Valla},
  \citenamefont {Pan}, \citenamefont {Gardner}, \citenamefont {Lee},\ and\
  \citenamefont {Chu}}]{Valla:2012jz}%
  \BibitemOpen
  \bibfield  {author} {\bibinfo {author} {\bibfnamefont {T.}~\bibnamefont
  {Valla}}, \bibinfo {author} {\bibfnamefont {Z.~H.}\ \bibnamefont {Pan}},
  \bibinfo {author} {\bibfnamefont {D.}~\bibnamefont {Gardner}}, \bibinfo
  {author} {\bibfnamefont {Y.~S.}\ \bibnamefont {Lee}}, \ and\ \bibinfo
  {author} {\bibfnamefont {S.}~\bibnamefont {Chu}},\ }\href@noop {} {\bibfield
  {journal} {\bibinfo  {journal} {Phys. Rev. Lett.}\ }\textbf {\bibinfo
  {volume} {108}},\ \bibinfo {pages} {117601} (\bibinfo {year}
  {2012})}\BibitemShut {NoStop}%
\bibitem [{\citenamefont {Scholz}\ \emph {et~al.}(2012)\citenamefont {Scholz},
  \citenamefont {S{\'a}nchez-Barriga}, \citenamefont {Marchenko}, \citenamefont
  {Varykhalov}, \citenamefont {Volykhov}, \citenamefont {Yashina},\ and\
  \citenamefont {Rader}}]{Scholz:2012kz}%
  \BibitemOpen
  \bibfield  {author} {\bibinfo {author} {\bibfnamefont {M.~R.}\ \bibnamefont
  {Scholz}}, \bibinfo {author} {\bibfnamefont {J.}~\bibnamefont
  {S{\'a}nchez-Barriga}}, \bibinfo {author} {\bibfnamefont {D.}~\bibnamefont
  {Marchenko}}, \bibinfo {author} {\bibfnamefont {A.}~\bibnamefont
  {Varykhalov}}, \bibinfo {author} {\bibfnamefont {A.}~\bibnamefont
  {Volykhov}}, \bibinfo {author} {\bibfnamefont {L.~V.}\ \bibnamefont
  {Yashina}}, \ and\ \bibinfo {author} {\bibfnamefont {O.}~\bibnamefont
  {Rader}},\ }\href@noop {} {\bibfield  {journal} {\bibinfo  {journal} {Phys.
  Rev. Lett.}\ }\textbf {\bibinfo {volume} {108}},\ \bibinfo {pages} {256810}
  (\bibinfo {year} {2012})}\BibitemShut {NoStop}%
\bibitem [{\citenamefont {Ishizaka}\ \emph {et~al.}(2011)\citenamefont
  {Ishizaka}, \citenamefont {Bahramy}, \citenamefont {Murakawa}, \citenamefont
  {Sakano}, \citenamefont {Shimojima}, \citenamefont {Sonobe}, \citenamefont
  {Koizumi}, \citenamefont {Shin}, \citenamefont {Miyahara},\ and\
  \citenamefont {Kimura}}]{Ishizaka:2011dy}%
  \BibitemOpen
  \bibfield  {author} {\bibinfo {author} {\bibfnamefont {K.}~\bibnamefont
  {Ishizaka}}, \bibinfo {author} {\bibfnamefont {M.~S.}\ \bibnamefont
  {Bahramy}}, \bibinfo {author} {\bibfnamefont {H.}~\bibnamefont {Murakawa}},
  \bibinfo {author} {\bibfnamefont {M.}~\bibnamefont {Sakano}}, \bibinfo
  {author} {\bibfnamefont {T.}~\bibnamefont {Shimojima}}, \bibinfo {author}
  {\bibfnamefont {T.}~\bibnamefont {Sonobe}}, \bibinfo {author} {\bibfnamefont
  {K.}~\bibnamefont {Koizumi}}, \bibinfo {author} {\bibfnamefont
  {S.}~\bibnamefont {Shin}}, \bibinfo {author} {\bibfnamefont {H.}~\bibnamefont
  {Miyahara}}, \ and\ \bibinfo {author} {\bibfnamefont {A.}~\bibnamefont
  {Kimura}},\ }\href@noop {} {\bibfield  {journal} {\bibinfo  {journal} {Nat.
  Mater.}\ }\textbf {\bibinfo {volume} {10}},\ \bibinfo {pages} {521} (\bibinfo
  {year} {2011})}\BibitemShut {NoStop}%
\bibitem [{\citenamefont {King}\ \emph {et~al.}(2011)\citenamefont {King},
  \citenamefont {Hatch}, \citenamefont {Bianchi}, \citenamefont {Ovsyannikov},
  \citenamefont {Lupulescu}, \citenamefont {Landolt}, \citenamefont {Slomski},
  \citenamefont {Dil}, \citenamefont {Guan}, \citenamefont {Mi}, \citenamefont
  {Rienks}, \citenamefont {Fink}, \citenamefont {Lindblad}, \citenamefont
  {Svensson}, \citenamefont {Bao}, \citenamefont {Balakrishnan}, \citenamefont
  {Iversen}, \citenamefont {Osterwalder}, \citenamefont {Eberhardt},
  \citenamefont {Baumberger},\ and\ \citenamefont {Hofmann}}]{King:2011bu}%
  \BibitemOpen
  \bibfield  {author} {\bibinfo {author} {\bibfnamefont {P.~D.~C.}\
  \bibnamefont {King}}, \bibinfo {author} {\bibfnamefont {R.~C.}\ \bibnamefont
  {Hatch}}, \bibinfo {author} {\bibfnamefont {M.}~\bibnamefont {Bianchi}},
  \bibinfo {author} {\bibfnamefont {R.}~\bibnamefont {Ovsyannikov}}, \bibinfo
  {author} {\bibfnamefont {C.}~\bibnamefont {Lupulescu}}, \bibinfo {author}
  {\bibfnamefont {G.}~\bibnamefont {Landolt}}, \bibinfo {author} {\bibfnamefont
  {B.}~\bibnamefont {Slomski}}, \bibinfo {author} {\bibfnamefont {J.~H.}\
  \bibnamefont {Dil}}, \bibinfo {author} {\bibfnamefont {D.}~\bibnamefont
  {Guan}}, \bibinfo {author} {\bibfnamefont {J.}~\bibnamefont {Mi}}, \bibinfo
  {author} {\bibfnamefont {E.~D.~L.}\ \bibnamefont {Rienks}}, \bibinfo {author}
  {\bibfnamefont {J.}~\bibnamefont {Fink}}, \bibinfo {author} {\bibfnamefont
  {A.}~\bibnamefont {Lindblad}}, \bibinfo {author} {\bibfnamefont
  {S.}~\bibnamefont {Svensson}}, \bibinfo {author} {\bibfnamefont
  {S.}~\bibnamefont {Bao}}, \bibinfo {author} {\bibfnamefont {G.}~\bibnamefont
  {Balakrishnan}}, \bibinfo {author} {\bibfnamefont {B.~B.}\ \bibnamefont
  {Iversen}}, \bibinfo {author} {\bibfnamefont {J.}~\bibnamefont
  {Osterwalder}}, \bibinfo {author} {\bibfnamefont {W.}~\bibnamefont
  {Eberhardt}}, \bibinfo {author} {\bibfnamefont {F.}~\bibnamefont
  {Baumberger}}, \ and\ \bibinfo {author} {\bibfnamefont {P.}~\bibnamefont
  {Hofmann}},\ }\href@noop {} {\bibfield  {journal} {\bibinfo  {journal} {Phys.
  Rev. Lett.}\ }\textbf {\bibinfo {volume} {107}} (\bibinfo {year}
  {2011})}\BibitemShut {NoStop}%
\bibitem [{\citenamefont {R{\"o}{\ss}ler}\ \emph {et~al.}(2014)\citenamefont
  {R{\"o}{\ss}ler}, \citenamefont {Jang}, \citenamefont {Kim}, \citenamefont
  {Tjeng}, \citenamefont {Fisk}, \citenamefont {Steglich},\ and\ \citenamefont
  {Wirth}}]{Rossler:2014kn}%
  \BibitemOpen
  \bibfield  {author} {\bibinfo {author} {\bibfnamefont {S.}~\bibnamefont
  {R{\"o}{\ss}ler}}, \bibinfo {author} {\bibfnamefont {T.~H.}\ \bibnamefont
  {Jang}}, \bibinfo {author} {\bibfnamefont {D.~J.}\ \bibnamefont {Kim}},
  \bibinfo {author} {\bibfnamefont {L.~H.}\ \bibnamefont {Tjeng}}, \bibinfo
  {author} {\bibfnamefont {Z.}~\bibnamefont {Fisk}}, \bibinfo {author}
  {\bibfnamefont {F.}~\bibnamefont {Steglich}}, \ and\ \bibinfo {author}
  {\bibfnamefont {S.}~\bibnamefont {Wirth}},\ }\href@noop {} {\bibfield
  {journal} {\bibinfo  {journal} {Proc. Natl. Acad. Sci.}\ }\textbf {\bibinfo
  {volume} {111}},\ \bibinfo {pages} {4798} (\bibinfo {year}
  {2014})}\BibitemShut {NoStop}%
\bibitem [{\citenamefont {Denlinger}\ \emph
  {et~al.}(2013{\natexlab{b}})\citenamefont {Denlinger}, \citenamefont {Allen},
  \citenamefont {Kang}, \citenamefont {Sun}, \citenamefont {Min}, \citenamefont
  {Kim},\ and\ \citenamefont {Fisk}}]{Denlinger13126636}%
  \BibitemOpen
  \bibfield  {author} {\bibinfo {author} {\bibfnamefont {J.~D.}\ \bibnamefont
  {Denlinger}}, \bibinfo {author} {\bibfnamefont {J.~W.}\ \bibnamefont
  {Allen}}, \bibinfo {author} {\bibfnamefont {J.-S.}\ \bibnamefont {Kang}},
  \bibinfo {author} {\bibfnamefont {K.}~\bibnamefont {Sun}}, \bibinfo {author}
  {\bibfnamefont {B.-I.}\ \bibnamefont {Min}}, \bibinfo {author} {\bibfnamefont
  {D.-J.}\ \bibnamefont {Kim}}, \ and\ \bibinfo {author} {\bibfnamefont
  {Z.}~\bibnamefont {Fisk}},\ }\href@noop {} {\bibfield  {journal} {\bibinfo
  {journal} {arXiv}\ } (\bibinfo {year} {2013}{\natexlab{b}})},\ \Eprint
  {http://arxiv.org/abs/1312.6636v2} {1312.6636v2} \BibitemShut {NoStop}%
\end{thebibliography}%

\newpage

\noindent {\bf Materials and Methods}

\setcounter{page}{1}
\noindent
\smhb  powder has been synthesized by borothermal reduction
of Sm$_2$O$_3$ with metallic B  powder under vacuum at 1900 K. The
powder was pressed to rods that were sintered under vacuum at 2000
K. The sintered rods of $\sim8$ mm diameter and 60 mm length then
have been used for single crystal growth with an inductive, crucible
free floating zone technique under 0.4 MPa Ar pressure.  The starting
components were amorphous natural B
and Sm$_2$O$_3$.  The purity was 99.9\%\ and 99.996\%\ in the case of B
and Sm$_2$O$_3$, respectively.  Single crystals with a diameter of
typically 6 mm and a length of up to 40 mm were grown using [100]
oriented seeds.  The growth procedure was repeated twice to remove
any porosity from the starting feed rods.  During the first passage
the crystallization rate was 1 mm/min which gave the primary crystal
without pores.  At the end of the first passage the melting zone
was frozen and the crystal was zone-melted again to the opposite
side with a crystallization rate of 0.4 mm/min while rotating the
crystal with 5 rpm.  Powder X-ray diffraction patterns of the crushed
\smhb crystal revealed the presence of the \smhb only.  Moreover,
Laue back scattering patterns from both ends of the crystal shows
a single crystal with [100] orientation.
The crystal employed for the present ARPES experiments showed a
residual resistance of 
$R(1 {\rm K}) / R(300 {\rm K})=5.3\times10^4$. 
Oriented slabs with
length up to 5 mm and typical cross section of $1\times1$ mm$^2$
have been cut for the ARPES experiments.  The orientation was again
verified using X-ray Laue diffraction which gave sharp reflections
and did not show any twinning.

Samples were cleaved in ultrahigh vacuum ($<10^{-10}$ mbar) at
temperatures below 40 K. The crystal is found to cleave with two
distinct terminations: An apparently purely B terminated surface
which is characterized by a complex B 1s spectrum, shown in Fig.~1(i),
and a relatively simple valence band spectrum which contains the
\smtwopl \fmult ($4f^6 \rightarrow 4f^5$ photoemission transition)
[Fig.~\ref{fig:G}(j)]. The other termination apparently exposes Sm
atoms exclusively, evidenced by the simple \Bos spectrum and the
additional surface components in the valence band. Similar variations
of the valence band spectrum have recently been observed
\cite{Denlinger13126636}.  All results presented in this work were
obtained with surfaces of which the size of regions with a single
termination exceeds the size of the synchrotron beam profile.
Regardless of the termination, we observe the effect of a potential
with twice the lattice constant, presumably due to several domains
of a $(2 \times 1)$ surface reconstruction reported in scanning
tunneling microscopy \cite{Rossler:2014kn}.

Photoemission experiments were performed with the $1^3$ ARPES
end-station at the UE112--PGM2b beamline of BESSY II. Data at \hv{31}
were obtained with an energy resolution of 3 meV. A sample temperature
of 1 K is used unless indicated otherwise. \kperp values are
calculated assuming a free electron final state using an inner
potential of 14 eV.

\end{document}